\documentclass[11pt]{article}

\pdfoutput=1

\usepackage{epsfig}
\usepackage{graphicx}

\usepackage{latexsym,times}
\usepackage{amsmath}
\usepackage{amsxtra}
\usepackage{amscd}
\usepackage{amsthm}
\usepackage{amsfonts}
\usepackage{amssymb}
\usepackage{dsfont}
\usepackage{xcolor}
\usepackage{cancel}
\usepackage{framed}

\usepackage{esint}

\usepackage{enumitem}
\usepackage{float}

\usepackage{hyperref}
\usepackage[square,sort&compress,comma,numbers]{natbib}

\makeatletter

\@addtoreset{equation}{section} \makeatother

\renewcommand{\arraystretch}{1.5}       
\newcommand{\half}{{{\textstyle\frac{1}{2}}}}
\newcommand{\quarter}{{{\textstyle\frac{1}{4}}}}
\newcommand{\be}{\begin{equation}}
\newcommand{\ee}{\end{equation} }
\newcommand{\beqa}{\begin{eqnarray} }
\newcommand{\eeqa}{\end{eqnarray} }
\newcommand{\ba}{\begin{array}}
\newcommand{\ea}{\end{array}}
\newcommand{\bpm}{\begin{pmatrix}}
\newcommand{\epm}{\end{pmatrix}}

\newcommand{\so}{\mathbf{so}}

\newcommand{\Spin}{\mathbf{Spin}}

\newcommand{\SL}{\mathbf{SL}}

\newcommand{\rmd}{{\rm d}}

\newcommand{\ODD}{\mathbf{O}(D,D)}

\newcommand{\SpinD}{{\Spin(1,D{-1})}}
\newcommand{\oSpinD}{{{\Spin}(D{-1},1)}}

\newcommand{\Off}{\mathbf{O}(4,4)}

\newcommand\rphoton{r_{\scriptscriptstyle{\rm photon}}}
\newcommand\Rphoton{R_{\scriptscriptstyle{\rm photon}}}

\newcommand{\DFT}{\rm{DFT}}

\newcommand\Tr{{\rm Tr}}

\newcommand\rd{{\rm d}}
\newcommand\rD{{\rm D}}

\newcommand{\Vo}{V_{\scriptscriptstyle{\rm{orbit}}}}

\newcommand\cA{{\cal A}}

\newcommand\cD{{\cal D}}

\newcommand\cF{{\cal F}}

\newcommand\cH{{\cal H}}

\newcommand\cJ{{\cal J}}

\newcommand\cL{{\cal L}}
\newcommand\cM{{\cal M}}

\newcommand\cQ{{\cal Q}}

\newcommand\hcL{{\hat{\cal L}}}

\newcommand\dis{\displaystyle}

\def\tx{\tilde{x}}

\def\brpsi{\bar{\psi}}

\def\brq{{\bar{q}}}

\def\brPhi{{{\bar{\Phi}}}}

\def\brP{{\bar{P}}}

\newcommand{\na}{{\nabla}}
\newcommand{\trd}{{\bigtriangledown}}

\newcommand{\red}[1]{{\color{red} #1 \color{black}}}

\newcommand{\blue}[1]{{\color{blue} #1 \color{black}}}

\newcommand\p\partial

\begin{document}

\begin{titlepage}
\title{\vskip -100pt
\vskip 2cm 
The rotation curve of a point particle in  stringy gravity}

\author{\sc Sung Moon Ko,${}^{\,\sharp}$\quad Jeong-Hyuck Park${}^{\,\sharp\,\flat}$\quad and \quad  Minwoo Suh${}^{\,\sharp}$}

\date{}
\maketitle 
\begin{center}
${}^{\sharp}$Department of Physics, Sogang University, 35 Baekbeom-ro, Mapo-gu,  Seoul  04107, Korea\\
${}^{\flat}$B. W. Lee Center for Fields, Gravity and Strings, Institute for Basic Science, Daejeon 34047, Korea\\
~\\
\texttt{sinsmk2003@sogang.ac.kr\qquad  park@sogang.ac.kr\qquad minsuh@usc.edu  }
~\\
~~~\\
\end{center}
\begin{abstract}
\noindent    Double Field Theory  suggests to  view  the whole massless sector of closed strings  as the gravitational unity. The  fundamental symmetries therein, including  the $\mathbf{O}(D,D)$ covariance,  can  determine  unambiguously        how  the  Standard Model as well as  a relativistic point particle  should couple to the closed string massless   sector. The theory also refines the  notion of singularity. We consider  the most general,   spherically symmetric, asymptotically flat, static
   vacuum solution to ${D=4}$  Double Field Theory, which  contains three  free parameters and consequently   
 generalizes     the Schwarzschild geometry.   Analyzing the    circular geodesic  of  a point particle in string frame, we  obtain the    orbital velocity  as a function of $R/(M_{\scriptscriptstyle{\infty}}G)$ which is the dimensionless radial variable normalized by mass. The rotation curve  generically  features a maximum and thus non-Keplerian over a  finite range, while becoming   asymptotically Keplerian at infinity, $R/(M_{\scriptscriptstyle{\infty}}G)\rightarrow \infty$.  The  adoption  of the string frame rather than Einstein frame is the consequence of   the fundamental symmetry principle. Our result opens up a  new scheme to  solve the dark matter/energy problems by modifying General Relativity at `short' range of $R/(M_{\scriptscriptstyle{\infty}}G)$.

\end{abstract} 
\thispagestyle{empty}
\end{titlepage}
\newpage
\tableofcontents 


\newpage

\section{Introduction}

\begin{figure}[H]
\centering\includegraphics[width=90mm]{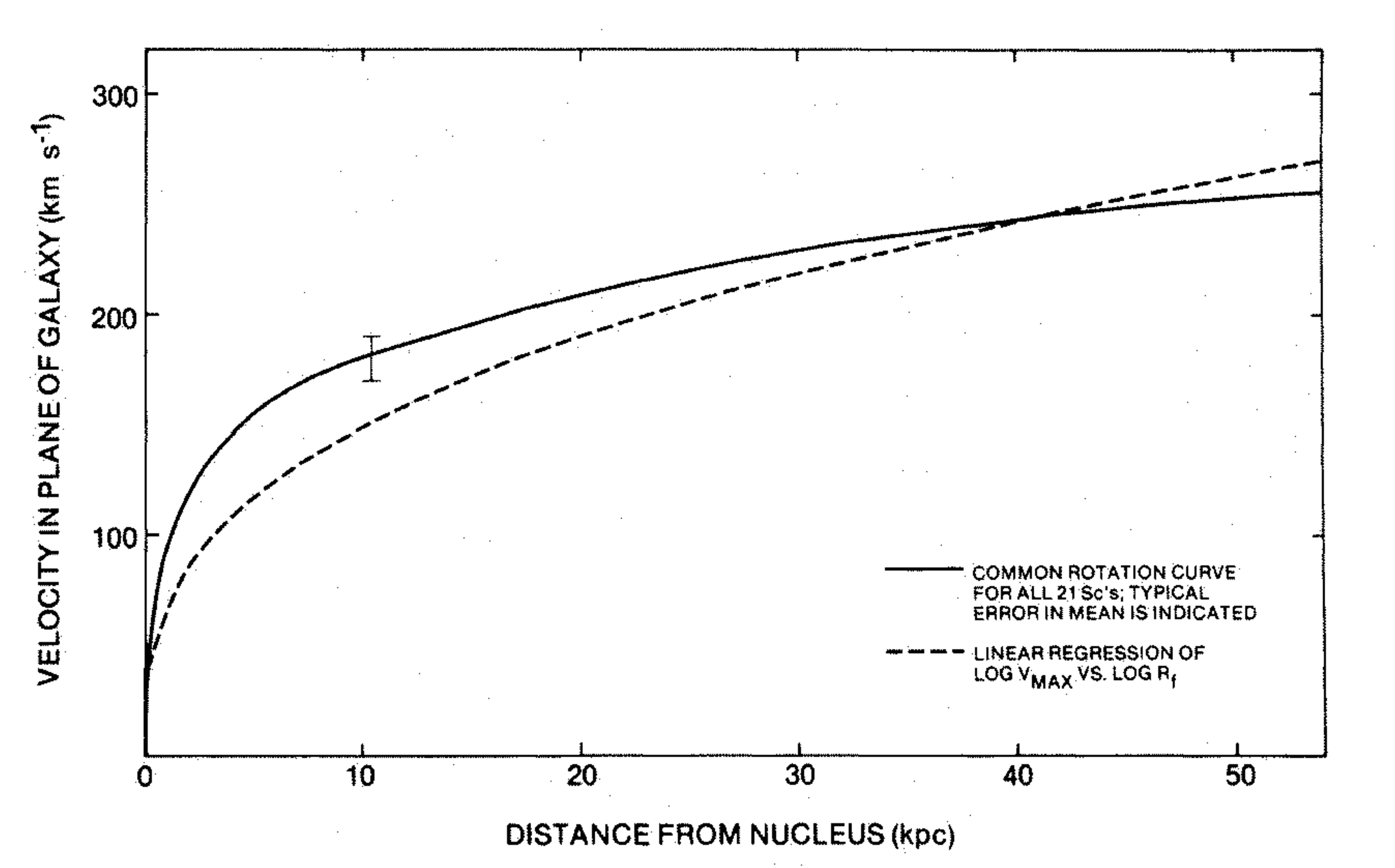}
\caption{{\bf{Observed galaxy rotation curve from  Ref.\cite{Rubin:1980zd} (FIG.~7 therein).}} 
The  curve   shows a fairly rapid  velocity rise 
and a slower rise (or flat) thereafter. For more figures, we refer  to \cite{Persic:1995ru,Sofue:2000jx}, or  \blue{\href{https://www.google.co.kr/search?q=galaxy+rotation+curve&newwindow=1&client=firefox-b&source=lnms&tbm=isch&sa=X&ved=0ahUKEwjc7tejvMfNAhXHp5QKHWGUAFIQ_AUICCgB&biw=1280&bih=604}{Google Search}. }}
\label{FIGObservation}
\end{figure}

The galaxy  rotation curve is a plot of the orbital velocities of visible stars versus their radial distance from the galactic  center,  see Figure~\ref{FIGObservation}. While Einstein gravity, \textit{i.e.~}General Relativity,  predicts the  Keplerian (inverse square root)  monotonic  fall-off  of the velocities,  observations  however      show  rather  `flat' ($100 \sim 200$ km/s)  curves after   a fairly rapid  rise~\cite{Rubin:1980zd}. The  resolution of the discrepancy might call for  {dark matter} or  modifications  of the law of gravity~\cite{Milgrom:1983ca}, or perhaps  both, \textit{e.g.~}\cite{Berezhiani:2015bqa}.   However -- despite  remarkable  improvements  of experimental sensitivity --  there has been no direct evidence of detecting any dark matter candidate.  This failure  might well motivate to explore various possibilities of modifying gravity,  General Relativity.

\noindent In General Relativity  the metric is the only geometric object. All other fields are viewed as matter or radiation; 
they source  the gravity. On the other hand, string theory puts a two-form gauge potential and a scalar dilaton   on an equal footing along with the metric, since the three of them,  conventionally denoted by $g_{\mu\nu}$, $B_{\mu\nu}$, $\phi$,  correspond to the massless sector of closed strings and  form a  multiplet of T-duality. This may indicate the existence of an alternative gravitational theory where the whole closed string massless  sector becomes geometric as  \textit{the gravitational unity}.  Such an idea, or Stringy Gravity,  has been materialized in recent years through the developments of   Double Field Theory (DFT).\\  

\noindent The primary goal of DFT~\cite{Siegel:1993xq,Siegel:1993th,Hull:2009mi} was to reformulate  supergravity with  doubled coordinates, $x^{A}=(\tx_{\mu},x^{\nu})$,  in a way that realizes    T-duality as a manifest symmetry of the  action and  unifies  diffeomorphisms and $B$-field gauge symmetry into `doubled diffeomorphisms'~\cite{Hull:2009zb,Hohm:2010jy,Hohm:2010pp}.        The closed string massless  sector should be then better represented by T-duality  or $\ODD$ covariant field variables, namely the DFT-metric, $\cH_{AB}$, and the  DFT-dilaton, $d$. The underlying differential geometry   has been subsequently   explored in various manners~\cite{Siegel:1993th,Jeon:2010rw,Hohm:2010xe,Jeon:2011cn,Jeon:2011vx,Hohm:2011si,Jeon:2012kd,Berman:2013uda,Cederwall:2014kxa} which   all suggested  to   generalize  the Riemannian geometry, often making contact with the `Generalized Geometry' \textit{a la} Hitchin~\cite{Hitchin:2004ut}, \textit{e.g.~}\cite{Coimbra:2011nw} (we refer  to review papers~\cite{Aldazabal:2013sca,Berman:2013eva,Hohm:2013bwa} on     various aspects of DFT).     \\

\noindent In particular,  in \cite{Jeon:2011cn} based on \cite{Jeon:2010rw}, the stringy  extension of the  Christoffel connection, `$\mathbf{\Gamma}_{A}$',  was derived~\eqref{DFTChristoffel}  which is made up of the whole closed string massless  sector now given by  $\cH_{AB}$ and $d$. Subsequently, it constitutes the  two-indexed (Ricci-type)   as well as zero-indexed (scalar)  covariant curvatures of DFT~\eqref{DFTcurvatures}, and hence furnishes the theory  with  geometrical interpretations. Further, (from the covariant constancy of the DFT-vielbeins),  the  connection, $\mathbf{\Gamma}_{A}$, determines  a pair of  spin connections, $\mathbf{\Phi}_{A}$ \&   $\mathbf{\brPhi}_{A}$, for the \textit{doubled} local Lorentz symmetries, $\SpinD\times\oSpinD$~\cite{Jeon:2011vx}. This  twofold  spin group    reflects the existing  two separate locally inertial frames for each left and right closed string mode~\cite{Duff:1986ne}.   Crucially, combining all the connections,  a master   derivative is at our disposal (see \cite{Choi:2015bga} for a concise review), 
\be
\mathbf{\cD}_{A}
=\mathbf{\partial}_{A}+\mathbf{\Gamma}_{A}+\mathbf{\Phi}_{A}
+\mathbf{\brPhi}_{A}\,,
\label{master}
\ee
which  takes care of the  fundamental symmetries of the  stringy gravity, \textit{i.e.~}DFT:
{\begin{itemize}
\setlength\itemsep{0em}
\item[--] \textit{$\ODD$ T-duality,}
\item[--] \textit{Doubled diffeomorphisms,}
\item[--] \textit{Twofold local Lorentz symmetries.}
\end{itemize}}
\noindent  The master derivative has been successfully utilized    to  complete the full order   supersymmetrizations of DFT~\cite{Jeon:2011sq,Jeon:2012hp,Cho:2015lha}, making each   term in every  formula      completely covariant under  the  fundamental symmetries   (\textit{c.f.~}\cite{Hohm:2011zr,Coimbra:2011nw}).\\

\noindent Besides the direct applications to  string theory,  the master derivative   naturally provides the \textit{minimal coupling} of  the closed string massless  sector to the Standard Model~\cite{Choi:2015bga}.  Each fermion therein  couples to the closed string massless  sector  as~\cite{Coimbra:2011nw,Jeon:2011vx,Choi:2015bga},
\footnote{Consequently, each fermion sources the two-indexed  DFT-curvature, $S_{p\brq}$, by $\brpsi\gamma_{p}\cD_{\brq}\psi$~\cite{Jeon:2011sq,Jeon:2012hp}. } 
\be
\ba{ll}
e^{-2d}\,\brpsi\gamma^{A}\cD_{A}\psi&=e^{-2d}\,\brpsi\gamma^{A}({\partial}_{A}\psi+\quarter\mathbf{\Phi}_{Apq}\gamma^{pq}\psi)\\
{}&\equiv\frac{1}{\sqrt{2}}\sqrt{-g}e^{-2\phi}\,\brpsi\gamma^{\mu} \left( \partial_{\mu} \psi + \frac{1}{4} \omega_{\mu  pq} \gamma^{pq} \psi + \frac{1}{24} H_{\mu pq} \gamma^{pq} \psi - \partial_{\mu} \phi \psi \right)\\
{}&\equiv\sqrt{-g}\,\bar{\chi}\gamma^{\mu} \left( \partial_{\mu} \chi + \frac{1}{4} \omega_{\mu  pq} \gamma^{pq} \chi + \frac{1}{24} H_{\mu pq} \gamma^{pq} \chi \right)\,,
\ea
\label{minimalcouplingF}
\ee
where  the $\Off$ covariant DFT-field variables on the top line  have been parametrized, (`$\equiv$'),   in terms of    the conventional (undoubled) spin connection,  $\omega_{\mu pq}$, the $H$-flux and  the scalar dilaton, $\phi$. Further, especially for the last expression, the field redefinition of the fermion, $\chi\equiv 2^{-\frac{1}{4}} e^{-\phi}\psi$,  has been performed to remove  the scalar dilaton completely.  The result of \eqref{minimalcouplingF} shows that,  (not only the fundamental   string but also) the Standard Model fermions  can  source the $H$-flux! It indicates the stringy nature,  if not  origin, of the fermion, $\chi$.  Similarly,  gauge bosons  in the Standard Model  couple to the closed string massless  sector as~\cite{Choi:2015bga,Jeon:2011kp}
\be
\ba{ll}
e^{-2d}\,\Tr\left(P^{AB}\brP^{CD}\cF_{AC}\cF_{BD}\right)
&\equiv-\frac{1}{4}\sqrt{-g}e^{-2\phi}\,\Tr\left(g^{\kappa\lambda}g^{\mu\nu}F_{\kappa\mu}F_{\lambda\nu} \right)\,,
\ea
\label{minimalcouplingB}
\ee
and hence they can source $\phi$, the scalar dilaton.  Surely, all the fields in the Standard Model source the (string framed) metric, $g_{\mu\nu}$.  This line of development  suggests that DFT is  not a mere reformulation of supergravity; it gives rise to  \textit{the stringy extension of   General Relativity}  as a (theoretically)  possible alternative theory of gravity.\\

\noindent It is the purpose of the present paper to push this idea further,  and to  derive  novel theoretical predictions which differ from that of Einstein gravity and  can be, in principle,  tested   against observations.  Yet, we intend   neither  to make any provocative  claim that DFT should replace General Relativity   as the correct theory of gravity, nor to  falsify DFT comparing with  precise  observational data.  Rather,   we merely  aspire  to postulate   DFT as a `theoretically' plausible gravitational theory and  to    explore its   various physical  aspects, especially the  implications  of the fundamental symmetries  of $\ODD$ T-duality and doubled diffeomorphisms.   \\

In the sense that  DFT does not take   only  the  metric, $g_{\mu\nu}$, as the   gravitational fields,  it is somewhat   similar to  the Brans-Dicke theory  where the gravitational interaction is meditated by a scalar field as well as the metric.  The Brans-Dicke theory contains a tunable dimensionless parameter, $\omega$, while  DFT  does not admit any free parameter as strongly constrained by   the  fundamental symmetries. Observations of  the  light deflection  in solar system -- in particular derived from the Cassini-Huygens experiment --  currently  set the lower bound, $\omega>40,000$. From the ``principle" of Occam's razor, the Brans-Dicke theory  appears  then  less favored  in comparison to   General Relativity.    This might motivate a haste  tendency to rule out   any gravitational theory   with a massless scalar field, such as DFT.  However,  because of the enlarged fundamental  symmetries,  DFT includes $H$-flux in addition to a scalar dilaton, $\phi$, and  consequently it  enriches the  possible spherically symmetric vacuum geometry -- see (\ref{SOL2}) -- and renders  \textit{a priori} more room to meet the observational constraints, \textit{e.g.~}light deflection.  Furthermore,  after field redefinition, $\phi\rightarrow\Phi:=e^{-\phi}$,   the scalar, $\Phi$, acquires an effective mass given by the scalar curvature in the Lagrangian~\eqref{conventionalNSNSaction}, as $4\partial_{\mu}\Phi\partial^{\mu}\Phi+R\Phi^{2}$. Thus, on a curved background, the scalar $\Phi$ is effectively massive and can modify the short distance gravity, which  is indeed the case as we show in the present work. Note that, for this, it is crucial to adopt  not the Einstein  frame but  the string frame.  We shall justify this choice of the frame from the symmetry principle.\\

\noindent In this work,    motivated by the Standard Model coupling  to the closed string massless  sector~\cite{Choi:2015bga}, \textit{e.g.~}(\ref{minimalcouplingF}), (\ref{minimalcouplingB}),   we firstly  look for   spherically symmetric vacuum solutions to  ${D=4}$  DFT,  which are in analogy to the    Schwarzschild solution to Einstein gravity.   We address   the notion of  spherical symmetry in DFT, in terms of \textit{a priori} not  the conventional field variables, $\{g_{\mu\nu},B_{\mu\nu},\phi\}$, but   the $\Off$ covariant DFT-metric and  DFT-dilaton, $\{\cH_{AB},d\}$: we spell  three DFT-Killing vectors~(\ref{Killing}) which  form an $\so(3)$ algebra through  C-bracket~(\ref{so3Hd}).  By solving directly   the DFT-Killing equations and  the Euler-Lagrangian equations of motion, we identify    the most  general form of the spherically symmetric,  asymptotically flat, static vacuum solutions to ${D=4}$  DFT, (\ref{THESOL0}), which turns  out to possess   three free parameters, including one for the  electric $H$-flux. \\

\noindent Though the backbone of the present work is the fundamental symmetry  principle of DFT, in practice,  with the spherically symmetric  ansatz~\eqref{ansatz},  we are solving  the full  Euler-Langrangian equations  of the rather familiar   gravity action of the closed string massless  sector~(\ref{conventionalNSNSaction}).   
They are equivalent to the vanishing of  both the two-indexed Ricci  and  the  zero-indexed  scalar DFT-curvatures~\eqref{DFTcurvatures}, and thus the solution can be identified as the spherically symmetric  \textit{DFT-vacuum}.  Essentially,  our result of the solution  is   a re-derivation of the known one by Burgess, Myers and Quevedo (BMQ)~\cite{Burgess:1994kq}. Historically,   Fischer  in 1948~\cite{Fisher:1948yn},  and    Janis, Newman and Winicour  later in 1968~\cite{Janis:1968zz} (F-JNW)   obtained  the most general spherical solution to    the Einstein gravity coupled to a massless scalar field. Then  adding one more scalar, or an \textit{axion}, and making use of the  $\SL(2,\mathbb{R})$ S-duality, BMQ managed to generate a   three-parameter family of spherical solutions. The axion is dual to the $H$-flux and our solution fully  agrees with the BMQ solution. Yet, since they focused on a time-independent dual scalar, the  possibility of having magnetic $H$-flux was excluded  from the beginning in their analysis. Our spherically symmetric ansatz allows both  electric and magnetic $H$-fluxes. Nonetheless we  show   that, the magnetic $H$-flux is inconsistent with the asymptotic flatness, see (\ref{eorm}) and (\ref{DEQ}).\\

\noindent Having more than one parameters,  the BMQ solution is generically `hairy'; the center   would  correspond to  a ``naked singularity''. Only if the scalar dilaton, $\phi$, and the $H$-flux (axion) are trivial, the solution is free of a naked singularity and  reduces identically  to the Schwarzschild metric. However, strictly speaking,  within the framework of DFT, the notion of  singularity should be addressed  in terms of its own  covariant curvatures.  Since we are solving for the \textit{DFT-vacuum} with the {vanishing  DFT-curvatures}  (both two-index and zero-index), while   there seems  no fully covariant  Riemann-type  four-index  curvature in DFT~\cite{Jeon:2011cn,Hohm:2011si},   
we shall rather not be concerned with  the issue of singularity (\textit{c.f.}~\cite{Virbhadra:2002ju} for an intriguing analysis on the photon sphere  of the  F-JNW geometry).\\

\noindent Once the spherically symmetric  vacuum solution  to  ${D=4}$ DFT is fully  identified, we shall proceed to analyze the geodesic  of a point particle on the vacuum geometry and derive the corresponding rotation curve, with the intention  of making a comparison with the galaxy observations [Figures~\ref{FIGObservation} and \ref{FIGgrcMAIN}]. In contrast to the null geodesic of a massless photon, the  massive  particle  geodesic   depends sensitively on the choice of the frame, \textit{i.e.~}string (Jordan) versus    Einstein.  Whilst     this ambiguity cannot be resolved to  full satisfaction in the conventional theories based on the Riemannian geometry (\textit{c.f.~}\cite{Faraoni:1998qx}), we show that the fundamental symmetries of DFT do the job: the symmetries of $\ODD$ T-duality  and doubled diffeomorphisms    dictate  that  the point particle  should follow  the geodesic defined not in the Einstein frame but in the string frame. Specifically,   we spell an $\ODD$ covariant (doubled)  action for a relativistic point particle coupled to the DFT-metric in (\ref{particleaction}) which can reduce consistently to the conventional (undoubled) particle action coupled to the string frame metric, (\ref{particleaction2}).\\

\noindent The rest of the present paper is organized as follows.  In section~\ref{SECparticleDFT},  we first review the concept  of `doubled-yet-gauged spacetime'~\cite{Park:2013mpa},  which provides a geometric  meaning to the doubled coordinates and the associated  section condition.   These two  are characteristics of  DFT, \textit{i.e.~}the stringy extension of Einstein gravity. We   spell an  action for a relativistic point particle  which propagates in the doubled-yet-gauged spacetime  and couples to the closed string massless  sector in an $\ODD$ covariant manner, (\ref{particleaction}). We also review briefly the geometric formulation of DFT and its Euler-Lagrangian equations of motion. 
Section~\ref{SECMAIN} contains most of our main results. We write the three  DFT-Killing vectors which form  the $\so(3)$ C-bracket relation. We identify the most general form of the   DFT-vacuum solutions which are spherically symmetric, static and asymptotically flat (\textit{c.f.~}BMQ~\cite{Burgess:1994kq}).  We then focus on  the circular geodesic motion of a relativistic point particle propagating around  the spherically symmetric DFT-vacuum. We compute the orbital  velocity and depict the  rotation curves as a function of radius  in various limits of the three free parameters. In contract to the Keplerian   (inverse square root) monotonic fall-off  on  the Schwarzschild geometry, the radial curve around a  generic,  spherically symmetric DFT-vacuum features a maximum. Yet, eventually at spatial infinity,  it becomes asymptotically Keplerian.  We conclude with comments   in section~\ref{SECCOM}. In particular, we    point out that, observations of galaxies far away might  well  reveal the short-distance nature of the gravitational law (\textit{c.f.~}`Cosmic Uroboros').  Appendix contains more rotation curves for various choices of the free parameters of the  DFT-vacua as well as some  technical derivations of the  main results.\\

\section{Point particle and stringy gravity in doubled-yet-gauged spacetime   \label{SECparticleDFT}}
\subsection{$\ODD$ covariant action for a point particle coupled to the DFT-metric}  
In order to describe the phenomenologically apparent,  four-dimensional spacetime, we employ the eight-dimensional,  \textit{douled-yet-gauged coordinate system}~\cite{Park:2013mpa}, where \textit{i)} an $\Off$ group is postulated with the $8\times 8$ invariant metric put in the off-block diagonal form, 
\be
\cJ_{AB}=
\left(\ba{cc}0&1\\1&0\ea\right)\,,
\ee
which along with its inverse, $\cJ^{AB}$, can freely lower or  raise the doubled vector indices, $A=1,2,\cdots, 8$, 
and \textit{ii)}  the doubled coordinates,  $x^{A}=(\tx_{\mu},x^{\nu})$, are gauged through  an equivalence relation,  namely  \textit{the coordinate gauge symmetry},
\be
x^{A}~\sim~x^{A}+\cJ^{AB}\Phi_{i}(x)\partial_{B}\Phi_{j}(x)\,.
\label{equivalence}
\ee
Here and henceforth, $\Phi_{i}$, $\Phi_{j}$ denote the arbitrary fields and their arbitrary derivatives which should  {`belong'} to the theory, \textit{i.e.~}DFT.  The  equivalence relation~(\ref{equivalence}) is realized  in DFT -- as for a target spacetime perspective --  simply by  requiring that all the functions are  invariant under the coordinate gauge symmetry shift,
\be
\ba{ll}
\Phi_{i}(x)=\Phi_{i}(x+\Delta)\,,\quad&\quad\Delta^{A}=
\Phi_{j}(x)\partial^{A}\Phi_{k}(x)\,.
\ea
\label{Delta}
\ee
In fact, as  can be seen easily from the power series expansion~\cite{Park:2013mpa,Lee:2013hma},  the invariance  is  equivalent to  the so-called  `{section condition}'~\cite{Siegel:1993th}:\footnote{
In \eqref{sectioncon}, the former (strong) constraint implies the latter (weak) one, since   $\,\partial_{A}\partial^{B}\Phi\,\partial_{B}\partial^{C}\Phi=0$ means that  $\partial_{A}\partial^{B}\Phi$ is a nilpotent matrix and hence is  traceless. Yet, replacing $\Phi_{k}$ by the product, $\Phi_{i}\Phi_{j}$,  the latter  may give the former.}
\be
\ba{ll}
\partial_{A}\Phi_{i}\partial^{A}\Phi_{j}=0\,,\quad&\quad
\partial_{A}\partial^{A}\Phi_{k}=0\,.
\ea
\label{sectioncon}
\ee
Upon the section condition, the generalized Lie derivatives~\cite{Siegel:1993th,Hull:2009zb}, 
\be
\hcL_{V}T_{M_{1}\cdots M_{n}}:=V^{N}\partial_{N}T_{M_{1}\cdots M_{n}}+\omega\partial_{N}V^{N}T_{M_{1}\cdots M_{n}}+\sum_{i=1}^{n}(\partial_{M_{i}}V_{N}-\partial_{N}V_{M_{i}})T_{M_{1}\cdots M_{i-1}}{}^{N}{}_{M_{i+1}\cdots  M_{n}}\,,
\label{tcL}
\ee
are  closed under commutations through so-called the C-bracket,
\be
\ba{cc}
\left[\hcL_{U},\hcL_{V}\right]=\hcL_{[U,V]_{\rm{C}}}\,,~~&~~~
{}[U,V]^{M}_{\rm{C}}:= U^{N}\partial_{N}V^{M}-V^{N}\partial_{N}U^{M}+\half V^{N}\partial^{M}U_{N}-\half U^{N}\partial^{M}V_{N}\,.
\ea
\ee 
Thus, it generates the diffeomorphisms in the doubled-yet-gauged spacetime.\\

\noindent On the other hand, on a particle worldline, or on a string worldsheet, the doubled coordinates are dynamical fields and need to be gauged  explicitly with the introduction of a relevant  gauge potential~\cite{Lee:2013hma},
\be
\rD x^{A}:=\rd x^{A}-\cA^{A}\,.
\ee
As in any gauge theory the gauge potential, $\cA^{A}$, should meet  precisely the same property as the gauge generator which is in the present  case,  `derivative-index-valued' $\Delta^{A}$ in (\ref{Delta}), such that
\be
\ba{ll}
\cA^{A}\partial_{A}=0\,,\quad&\quad \cA^{A}\cA_{A}=0\,.
\ea
\label{sectionconA}
\ee
It is crucial to note that $\rD x^{A}$ is a covariant  vector of DFT and also  is invariant under the coordinate gauge symmetry,  but  the ordinary infinitesimal one-form,  $\rd x^{A}$, is   anomalous~\cite{Lee:2013hma}.  Under infinitesimal diffeomorphisms, $\delta_{\scriptscriptstyle{V}} x^{A}=V^{A}(x)$, as well as the coordinate gauge symmetry, $\delta_{\scriptscriptstyle{\Delta}} x^{A}=\Delta^{A}(x)$,  provided  the the potential transforms   properly,  respecting (\ref{sectionconA}),  as
\be
\ba{ll}
\delta_{\scriptscriptstyle{V}} \cA^{A}
=-\partial^{A}V_{B}\cA^{B}+\partial^{A}V_{B}\,\rd x^{B}\,,\quad&\quad \delta_{\scriptscriptstyle{\Delta}} \cA^{A}=\rd\Delta^{A}\,,
\ea
\ee
we have the covariance as well as the invariance,
\be{
\ba{ll}  
\delta_{\scriptscriptstyle{V}}(\partial_{A})
=(\partial^{B}V_{A}-\partial_{A}V^{B})\partial_{B}\,,\quad&\quad
\delta_{\scriptscriptstyle{\Delta}} (\partial_{A})=0\,,\\
\delta_{\scriptscriptstyle{V}} (\rD  x^{A})=(\partial_{B}V^{A}-\partial^{A}V_{B})\rD x^{B}\,,\quad&\quad
\delta_{\scriptscriptstyle{\Delta}} (\rD  x^{A})=0\,.
\ea}
\ee
~\\
The fundamental symmetries, together with the coordinate gauge symmetry, then    uniquely fix the relativistic  point  particle action on a generic closed string  background: 
\be\dis{
S_{\scriptscriptstyle{\rm{particle}}}=
\int\rd\tau~e^{-1\,}\rD_{\tau}x^{A}\rD_{\tau}x^{B}\cH_{AB}(x)-\quarter m^{2}e\,,}
\label{particleaction}
\ee
where $e$ is an einbein, $m$ is the mass of the particle and $\cH_{AB}$ is the DFT-metric which is,  by definition,   a symmetric $\ODD$ element:
\be
\ba{ll}
\cH_{AB}=\cH_{BA}\,,\quad&\quad\cH_{A}{}^{C}\cH_{B}{}^{D}\cJ_{CD}=\cJ_{AB}\,.
\ea
\label{GM}
\ee
In general, up to $\ODD$ rotations, 
the section conditions of \eqref{sectioncon} and \eqref{sectionconA} are solved by letting
\be
\ba{ll}
\partial_{A}={(\tilde{\partial}^{\mu}\,,\,\partial_{\nu})}\equiv
(0\,,\,\partial_{\nu})\,,\quad&\quad
\cA_{A}\equiv(0\,,\,A_{\nu})\,.
\ea
\ee
Consequently, only the dual tilde-coordinates are gauged:  
\be
\rD_{\tau}x^{A}\equiv\left(\dot{\tilde{x}}_{\mu}-A_{\mu}\,,\,\dot{x}^{\nu}\right)\,.
\ee
Also, the  DFT-metric  and the DFT-dilaton can  be   conventionally parametrized  by  the  string frame metric,   the  $B$-field and the scalar   dilaton:\footnote{In fact, Eq.\eqref{NSNSparametrization} gives the generic parametrization of the DFT-metric whose  upper left $D\times D$ block is non-degenerate. If it is degenerate, the DFT-metric should be parametrized differently. Such  a background was explicitly obtained through  a  T-duality rotation along  temporal directions~\cite{Lee:2013hma}, and was shown in  \cite{Ko:2015rha} to  realize a   `non-relativistic' string background. }
\be
\ba{ll}
\cH_{AB}\equiv
\scriptstyle{\left(\ba{cc}{{g^{-1}}}&{{-g^{-1}B}}\\{{Bg^{-1}}}&{{\,\,g-Bg^{-1}B}}\ea\right)}\,,\quad&\quad e^{-2d}\equiv\sqrt{-g}e^{-2\phi}\,.
\ea
\label{NSNSparametrization} 
\ee 
An instructive  relation follows
\be
\rD_{\tau}x^{A}\rD_{\tau}x^{B}\cH_{AB}
\equiv\dot{x}^{\mu}\dot{x}^{\nu}g_{\mu\nu}+
\left(\dot{\tx}_{\mu}-A_{\mu}\!+\dot{x}^{\rho}B_{\rho\mu}\right)
\left(\dot{\tx}_{\nu}-A_{\nu}\!+\dot{x}^{\sigma}B_{\sigma\nu}\right)
g^{\mu\nu}\,.
\ee
Now, integrating out the auxiliary  gauge potential, $A_{\mu}$,  the fully symmetric  action~(\ref{particleaction}) reduces to the standard action for a relativistic point particle coupled only  to the string frame metric:
\be\dis{
S_{\scriptscriptstyle{\rm{particle}}}\equiv
\int\rd\tau~e^{-1\,}\dot{x}^{\mu}\dot{x}^{\nu}g_{\mu\nu}-\quarter m^{2}e\,.}
\label{particleaction2}
\ee
This implies that, the particle follows   the geodesic path  defined  in the \textit{string frame.}  We stress that    this preferred choice of the frame is due to the fundamental symmetry  principle of DFT.\\

\subsection{Pure DFT and its Euler-Lagrangian equations of motion}

In DFT, the  massless sector of closed strings  is represented by the DFT-dilaton, $d$,  and the DFT-metric, $\cH_{AB}$, satisfying the defining property~\eqref{GM}. With the $\ODD$ invariant metric, $\cJ_{AB}$, the latter defines 
 a pair of projectors,
\be
\ba{llll}
P_{AB}=\half(\cJ_{AB}+\cH_{AB})\,,&\quad P_{A}{}^{B}P_{B}{}^{C}=P_{A}{}^{C}\,,&\quad
\brP_{AB}=\half(\cJ_{AB}-\cH_{AB})\,,&\quad \brP_{A}{}^{B}\brP_{B}{}^{C}=\brP_{A}{}^{C}\,,
\ea
\ee
which are symmetric, orthogonal and complete,
\be
\ba{llll}
P_{AB}=P_{BA}\,,\quad&\qquad
\brP_{AB}=\brP_{BA}\,,\quad&\qquad
P_{A}{}^{B}\brP_{B}{}^{C}=0\,,\quad&\qquad
P_{AB}+\brP_{AB}=\cJ_{AB}\,.
\ea
\ee
The  stringy  or DFT extension of the Christoffel connection  is,  from \cite{Jeon:2011cn},
\be
\ba{lll}
\Gamma_{CAB}&=&2\left(P\partial_{C}P\brP\right)_{[AB]}
+2\left({{\brP}_{[A}{}^{D}{\brP}_{B]}{}^{E}}-{P_{[A}{}^{D}P_{B]}{}^{E}}\right)\partial_{D}P_{EC}\\
{}&{}&\,-\,\textstyle{\frac{4}{D-1}}\left(\brP_{C[A}\brP_{B]}{}^{D}+P_{C[A}P_{B]}{}^{D}\right)\!\left(\partial_{D}d+(P\partial^{E}P\brP)_{[ED]}\right)\,.
\ea
\label{DFTChristoffel}
\ee
Further, if we set 
\be
R_{CDAB}:=\partial_{A}\Gamma_{BCD}-\partial_{B}\Gamma_{ACD}+\Gamma_{AC}{}^{E}\Gamma_{BED}-\Gamma_{BC}{}^{E}\Gamma_{AED}\,,
\ee
we may define  so-called the `semi-covariant' four-indexed Riemann-type DFT-curvature, 
\be
S_{ABCD}:=\half(R_{ABCD}+R_{CDAB}-\Gamma^{E}{}_{AB}\Gamma_{ECD})\,,
\ee
which in turn  sets  the completely covariant,   zero-indexed scalar  and two-indexed Ricci-type  DFT-curvatures,
\be
\ba{ll}
(P^{AB}P^{CD}-\brP^{AB}\brP^{CD})S_{ACBD}\,,\quad~~&~~\quad
P_{A}{}^{C}\brP_{B}{}^{D}S_{CED}{}^{E}\,.
\ea
\label{DFTcurvatures}
\ee
The scalar curvature defines the  pure DFT Lagrangian, multiplied by   the weightful DFT-dilaton factor, 
\be
\cL_{\DFT}=e^{-2d}(P^{AB}P^{CD}-\brP^{AB}\brP^{CD})S_{ACBD}\,.
\ee
The full Euler-Lagrangian  equations  of motion are  nothing but the vanishing of the two DFT-curvatures~\eqref{DFTcurvatures}.
\\

\noindent With the `conventional'  parametrization of the DFT-dilaton and the DFT-metric~\eqref{NSNSparametrization}, all the Euler-Lagrangian  equations  of  the pure DFT reduce to
\begin{eqnarray}
&&R_{\mu\nu}+2\trd_{\mu}\partial_{\nu}\phi-\quarter H_{\mu\rho\sigma}H_{\nu}{}^{\rho\sigma}=0\,,\label{EOMg}\\
&&\trd^{\lambda}H_{\lambda\mu\nu}
-2(\partial^{\lambda}\phi)H_{\lambda\mu\nu}=0\,,\label{EOMB}\\
&&R+4\Box\phi-4\partial_{\mu}\phi\partial^{\mu}\phi-\textstyle{\frac{1}{12}}H_{\mu\nu\rho}H^{\mu\nu\rho}=0\,.\label{EOMphi}
\end{eqnarray}
Basically,  the first two equations correspond to the symmetric and the anti-symmetric parts of the two-indexed DFT-curvature (after  pulled back by DFT-vielbeins, $S_{p\brq}$), while the last is precisely the scalar DFT-curvature. These formulae  can be also derived as the equations of motion of the conventional (undoubled) action for the  closed string massless  sector,
\be
\int\rd x^{D}~\sqrt{-g}e^{-2\phi}\Big(\,R+4\partial_{\mu}\phi\partial^{\mu}\phi
-\textstyle{\frac{1}{12}}H_{\mu\nu\rho}H^{\mu\nu\rho}\,\Big)\,.
\label{conventionalNSNSaction}
\ee
Eq.(\ref{EOMB}) can be rewritten in terms of the form notation,
\be
\rd\star\left(e^{-2\phi}H_{\scriptscriptstyle{(3)}}\right)=0\,.
\label{EOMB2}
\ee
Combining (\ref{EOMphi}) with the trace of (\ref{EOMg}), we have
\be
\Box\phi-2\partial_{\mu}\phi\partial^{\mu}\phi+
\textstyle{\frac{1}{12}}H_{\mu\nu\rho}H^{\mu\nu\rho}=0\,.
\label{EOMphi2}
\ee
After all, the Euler-Lagrangian equations of motion boil down to  (\ref{EOMg}), (\ref{EOMB2}) and (\ref{EOMphi2}).\\

It is worth while to note that the equations of motion ensure the  conservation of the Noether current,
\be
\ba{ll}
\trd_{\mu}J^{\mu}=0\,,\quad&\quad J^{\mu}=e^{-2\phi}\left(\partial^{\mu}\phi+\quarter H^{\mu\nu\rho}B_{\nu\rho}\right)\,,
\ea
\ee
which corresponds to the global scale symmetry present in the action~(\ref{conventionalNSNSaction}), 
\be
\ba{lll}
\phi~~~\rightarrow~~~\phi+(D-2)\lambda\,,\quad&\quad
g_{\mu\nu}~~~\rightarrow~~~e^{4\lambda}g_{\mu\nu}\,,\quad&\quad
B_{\mu\nu}~~~\rightarrow~~~e^{4\lambda}B_{\mu\nu}\,.
\ea
\ee

\newpage

\section{Circular geodesic  around  the spherical  vacuum in ${D=4}$ DFT\label{SECMAIN}}

In this section we spell our main results. Appendix contains detailed derivations and more rotation curves.

\begin{itemize}

\item \textbf{$\ODD$ covariant  action for a point particle in  doubled-yet-gauged spacetime}

We recall (\ref{particleaction}) that,   requiring  $\ODD$ T-duality, doubled diffeomorphisms and   the coordinate gauge symmetry,   the   action for a point  particle   in the doubled-yet-gauged spacetime is uniquely determined: 
\be\dis{
S_{\scriptscriptstyle{\rm{particle}}}=
\int\rd\tau~e^{-1\,}\rD_{\tau}x^{A}\rD_{\tau}x^{B}\cH_{AB}(x)-\quarter m^{2}e\,,}
\label{particleactionRecall}
\ee
which reduces to the conventional (undoubled)  action (\ref{particleaction2}) for a relativistic point particle coupled   to the string frame metric:
\be\dis{
S_{\scriptscriptstyle{\rm{particle}}}\equiv
\int\rd\tau~e^{-1\,}\dot{x}^{\mu}\dot{x}^{\nu}g_{\mu\nu}-\quarter m^{2}e\,.}
\label{particleaction2Recall}
\ee
Thus,  the  geodesic motion of the particle should be analyzed in the \textit{string frame.}  \\

\item \textbf{Spherically  symmetric ansatz for DFT}

We prescribe that any spherically symmetric DFT configuration should  admit three doubled Killing vectors,  $V^{A}_{a}$, $a=1,2,3$, which  satisfy the DFT-Killing equations  in terms of the generalized Lie derivative~\cite{Park:2015bza},
\be
\ba{ll}
\hcL{}_{V_{a}}{\cH_{AB}=0}\,,\quad&\quad
\hcL{}_{V_{a}}\!\left(e^{-2d}\right)\!=0\,,
\ea
\label{so3Hd0}
\ee
and form an  $\so(3)$ algebra through the $\mathbf{C}$-bracket,
\be
\left[V_{a},V_{b}\right]_{\mathbf{C}}=\sum_{c}\epsilon_{abc}V_{c}\,.
\label{so3Hd}
\ee
With (\ref{NSNSparametrization}),  
such a  spherically symmetric and  static closed string  background assumes the   generic form:
\be
\ba{l}
\rd s^{2}=e^{2\phi(r)}\left[-A(r)\rd t^{2} +A^{-1}(r)\rd r^{2}+A^{-1}(r)C(r)\,\rd\Omega^{2}\right],\\
B_{\scriptscriptstyle{(2)}}=
B(r)\cos\vartheta\,\rd r\wedge\rd\varphi+ h\cos\vartheta\,\rd t\wedge\rd\varphi\,,
\ea
\label{ansatz}
\ee
which contains  four   unknown radial  functions, $A(r)$, $B(r)$, $C(r)$ and the scalar dilaton, $\phi(r)$. We also set   $\rd\Omega^{2}=\rd\vartheta^{2}+\sin^{2}\vartheta\rd\varphi^{2}$ and put the $B$-field into a two-form, 
$B_{\scriptscriptstyle{(2)}}=\half B_{\mu\nu}\rd x^{\mu}\wedge\rd x^{\nu}$. The $H$-flux then takes the most general spherically symmetric form,
\[
H_{\scriptscriptstyle{(3)}}=\rd B_{\scriptscriptstyle{(2)}}=
B(r)\sin\vartheta\,\rd r\wedge\rd\vartheta\wedge\rd\varphi+ h\sin\vartheta\,\rd t\wedge\rd\vartheta\wedge\rd\varphi\,,
\label{Hflux}
\]
which is closed for constant $h$.

Writing $V^{A}_{a}=\left(\lambda_{a\mu},\xi_{a}^{\nu}\right)$, 
the doubled  $\so(3)$ Killing vectors are given concretely  by
\be\textstyle{
\ba{ll}
\!\!\lambda_{1}=\textstyle{\frac{\cos\varphi}{\sin\vartheta}}\big[h\rd t+B(r)\rd r\big]\,,&\xi_{1}=\sin\varphi\partial_{\vartheta}
+\cot\vartheta\cos\varphi\partial_{\varphi}\,,\\
\!\!\lambda_{2}=\textstyle{\frac{\sin\varphi}{\sin\vartheta}}\big[h\rd t+B(r)\rd r\big]\,,&\xi_{2}=-\cos\varphi\partial_{\vartheta}+
\cot\vartheta\sin\varphi\partial_{\varphi}\,,\\
\!\!\lambda_{3}=0\,,&\xi_{3}=-\partial_{\varphi}\,.
\ea}
\label{Killing}
\ee
They meet, with the ordinary (undoubled)  Lie derivative,  
\be
\ba{lll}
\cL_{\xi_{a}}g_{\mu\nu}=0\,,\quad&\quad
\cL_{\xi_{a}}\phi=\xi^{\mu}_{a}\partial_{\mu}\phi=0\,,
\quad&\quad
\cL_{\xi_{a}}B_{\scriptscriptstyle{(2)}}=-\rd{\lambda_{a}}\,,
\ea
\ee 
and  hence, as expected for the $H$-flux,  
\be
\cL_{\xi_{a}}H_{\scriptscriptstyle{(3)}}=0\,.
\ee

\item \textbf{Spherically symmetric, static and asymptotically flat vacuum solution to $D=4$ DFT}

We  insert the spherically symmetric static ansatz~(\ref{ansatz}) into the Euler-Lagrangian equations of motion, especially (\ref{EOMg}), (\ref{EOMB2}) and (\ref{EOMphi2}). We impose  the boundary condition of  the asymptotic flatness and obtain the most general form of such solutions.  Appendix~\ref{APPSOL} contains the details  of our  direct derivation of the solution.    The asymptotic flatness turns out to be inconsistent with the magnetic $H$-flux, and hence we put  ${B(r)=0}$ and 
\be
H_{\scriptscriptstyle{(3)}}= h\sin\vartheta\,\rd t\wedge\rd\vartheta\wedge\rd\varphi\,.
\ee  
{The most general, spherically symmetric, asymptotically flat, static vacuum solution  to ${D=4}$ DFT is, with the ansatz~(\ref{ansatz}),  given by}
\be\textstyle{
\ba{ll}
A(r)=\left(\frac{r-\alpha}{r+\beta}\right)^{\frac{a}{\sqrt{a^{2}+b^{2}}}}\,,~&C(r)=(r-\alpha)(r+\beta)\,,\\
\multicolumn{2}{c}{
B_{\scriptscriptstyle{(2)}}=h\cos\vartheta\,\rd t\wedge\rd\varphi\,,}\\
\multicolumn{2}{c}{e^{2\phi}=\gamma_{+}\left(\frac{r-\alpha}{r+\beta}\right)^{\frac{b}{\sqrt{a^{2}+b^{2}}}}+\gamma_{-}\left(\frac{r-\alpha}{r+\beta}\right)^{\frac{-b}{\sqrt{a^{2}+b^{2}}}}\,,}
\ea}
\label{THESOL0}
\ee
where $a,b,h$ are three real  constants satisfying $\,b^{2}\geq h^{2}$,  and $\alpha,\beta,\gamma_{\pm}$ are associated  shorthand,
\be
\ba{lll}
\alpha:=\frac{a}{a+b}\sqrt{a^{2}+b^{2}}\,,\quad&\quad
\beta:=\frac{b}{a+b}\sqrt{a^{2}+b^{2}}\,,\quad&\quad
\gamma_{\pm}:=\half(1\pm\sqrt{1-{h^{2}/b^{2}}})\,.
\ea
\label{constantsabg}
\ee
This result is a re-derivation of the  BMQ solution which was generated by the $\SL(2,\mathbb{R})$ S-duality~\cite{Burgess:1994kq}.  After shifting the radius, $r\rightarrow r-\beta$, we may rewrite the solution as
\be
{\ba{cc}
e^{2\phi}=\gamma_{+}
\left(1-\frac{\sqrt{a^{2}+b^{2}}}{r}\right)^{\frac{b}{\sqrt{a^{2}+b^{2}}}}
+\gamma_{-}
\left(1-\frac{\sqrt{a^{2}+b^{2}}}{r}\right)^{\frac{-b}{\sqrt{a^{2}+b^{2}}}}\,,\quad&\quad
B_{\scriptscriptstyle{(2)}}=h\cos\vartheta\,\rd t\wedge\rd\varphi\,,\\
\multicolumn{2}{c}{
\rd s^{2}=e^{2\phi}\left[-\left(1-\frac{\sqrt{a^{2}+b^{2}}}{r}\right)^{\frac{a}{\sqrt{a^{2}+b^{2}}}}\rd t^{2}
+\left(1-\frac{\sqrt{a^{2}+b^{2}}}{r}\right)^{\frac{-a}{\sqrt{a^{2}+b^{2}}}}\left(\rd r^{2}+r\big(r-\sqrt{a^{2}+b^{2}}\big)
\rd\Omega^{2}\right)\right],}
\ea}
\label{SOL2}
\ee
where the radial origin, $r=0$, corresponds to the coordinate singularity.

It is worth while to note the  expression of the DFT integral measure,
\be
e^{-2d}=e^{2\phi}CA^{-1}\sin\vartheta=g_{\vartheta\vartheta}(r)\sin\vartheta \,.
\ee

\item \textbf{Circular geodesic and  orbital velocity}

We proceed to analyze    the  circular geodesic, with both  $r$ and  $\vartheta\equiv\frac{\pi}{2}$ fixed.   
We  introduce   the `proper' radius,
\be
R:=\sqrt{g_{\vartheta\vartheta}(r)}=\sqrt{C(r)/A(r)}\,e^{\phi(r)}\,,
\label{properR0}
\ee
which would convert the   metric   into a  canonical form where the    angular part is  `properly' normalized (hence comparable to observations, \textit{e.g.~}galaxy rotation curves):
\be\textstyle{
\rd s^{2} =g_{tt}\rd t^{2}+g_{\scriptscriptstyle{RR}}\rd R^{2}+R^{2}\rd\Omega^{2}
=-e^{2\phi}A\,\rd t^{2} +e^{2\phi}A^{-1}
\left(\frac{\rd R}{\rd r}\right)^{-2}\rd R^{2}+R^{2}\rd\Omega^{2}\,.}
\label{canonicalmetric}
\ee
The  radial component of the  geodesic equation  determines the angular velocity as a function of  $r$,  or the proper radius, $R$,
\be
\textstyle{\left(\frac{\rd\varphi}{\rd t}\right)^{\!2}
={\frac{\rd~}{\rd r}(Ae^{2\phi})}\left/{\frac{\rd~}{\rd r}(CA^{-1}e^{2\phi})}\right.=-\half R^{-1}\frac{\rd g_{tt}}{\rd R}\,.}
\label{varthetat0}
\ee
The orbital velocity is  given by  the proper radius times the  angular velocity,  computable   from  \eqref{THESOL0}, \eqref{properR0}, \eqref{varthetat0},
\be
\textstyle{
\Vo=\left|R\frac{\rd\varphi}{\rd t}\right|=\left[-\half R\frac{\,\rd g_{tt}}{\rd R}\,\right]^{\frac{1}{2}}\,.}
\label{VorbitRg}
\ee
Clearly, from (\ref{varthetat0}) and (\ref{VorbitRg}), the gravitational force can be \textit{repulsive} if and only if the orbital velocity is pure imaginary, \textit{i.e.~}$\frac{\rd g_{tt}}{\rd R}>0$.  \\

\noindent Explicitly we have for the solution~\eqref{THESOL0},
\begin{eqnarray}
&&\!\!\!\!\!\!\!\!\!\!\!\!\!\!\!\!\!\!\!\!\!\!\!\!\!R=\textstyle{\left[(r-\alpha)(r+\beta)\left(\gamma_{+}\left(\frac{r-\alpha}{r+\beta}\right)^{\frac{-a+b}{\sqrt{a^{2}+b^{2}}}}+
\gamma_{-}\left(\frac{r-\alpha}{r+\beta}\right)^{\frac{-a-b}{\sqrt{a^{2}+b^{2}}}}\right)\right]^{\frac{1}{2}}\,,}\label{explicitR}\\
&&\!\!\!\!\!\!\!\!\!\!\!\!\!\!\!\!\!\!\!\!\!\!\!\!\!\left(\frac{\rd\varphi}{\rd t}\right)^{2}=
\textstyle{{\frac{1}{
\left(r-\alpha\right)
\left(r+\beta\right)}}
\left(\frac{r-\alpha}{r+\beta}\right)^{\frac{2a}{\sqrt{a^{2}+b^{2}}}}
\left[\frac{\gamma_{+}(a+b)\left(\frac{r-\alpha}{r+\beta}\right)^{\frac{2b}{\sqrt{a^{2}+b^{2}}}}~+\,\gamma_{-}(a-b)
}{\gamma_{+}
\left(2r-\alpha+\beta-a+b\right)\left(\frac{r-\alpha}{r+\beta}\right)^{\frac{2b}{\sqrt{a^{2}+b^{2}}}}~
+\,\gamma_{-}\left(2r-\alpha+\beta-a-b\right)}
\right]\,,}\\
&&\!\!\!\!\!\!\!\!\!\!\!\!\!\!\!\!\!\!\!\!\!\!\!\!\!\Vo=\textstyle{
\left[
\frac{\left(\gamma_{+}\left(\frac{r-\alpha}{r+\beta}\right)^{\frac{a+b}{\sqrt{a^{2}+b^{2}}}}+
\gamma_{-}\left(\frac{r-\alpha}{r+\beta}\right)^{\frac{a-b}{\sqrt{a^{2}+b^{2}}}}\right)\left(\gamma_{+}(a+b)\left(\frac{r-\alpha}{r+\beta}\right)^{\frac{2b}{\sqrt{a^{2}+b^{2}}}}+\gamma_{-}(a-b)\right)
}{\gamma_{+}
\left(2r-\alpha+\beta-a+b\right)\left(\frac{r-\alpha}{r+\beta}\right)^{\frac{2b}{\sqrt{a^{2}+b^{2}}}}
+\gamma_{-}\left(2r-\alpha+\beta-a-b\right)}
\right]^{\frac{1}{2}}
\,.}
\label{RvarphiV}
\end{eqnarray}
We refer readers to Appendix~\ref{APPGEO} for  details.

\item \textbf{Physical observables of the spherically symmetric DFT-vacuum}

There are   three  \textit{physical observables},\footnote{Alternative to \eqref{observables}, by analyzing the asymptotic behaviors of the scalar dilaton $\phi$ and the $H$-flux, one may obtain the dilaton and the $H$-flux  ``charges''~\cite{Burgess:1994kq}. But, since they are merely  parametrization-dependent  components of the DFT-metric and the DFT-dilaton,   the  ``charges'' cannot   quite be qualified as    $\ODD$ covariant quantities,  nor physical observables.  } on account of  the  three  free parameters, $a,b,h$ (${b^{2}\geq h^{2}}$),
\be
\textstyle{
\ba{l}
M_{\scriptscriptstyle{\infty}}G:={\dis{\lim_{R\rightarrow\infty}} }(R\Vo^{2})=\frac{1}{2}(a+b\sqrt{1-h^{2}/b^{2}})\,,\\
\Rphoton=R(\rphoton)\,,~\quad\rphoton=a+\half\!\left(\frac{a-b}{a+b}\right)\sqrt{a^{2}+b^{2}}\,,\\
\cQ[\partial_{t}]=\frac{1}{4}\left[a+\left(\frac{a-b}{a+b}\right)\sqrt{a^{2}+b^{2}}\,\right]\,.
\ea}
\label{observables}
\ee
The first defines   the asymptotic mass, $M_{\scriptscriptstyle{\infty}}$, from the  Keplerian fall-off of the orbital speed which, from (\ref{explicitR}), (\ref{RvarphiV}), eventually  takes place    at spatial infinity,
\be
\ba{lll}
\left.\rd s^{2}\right|_{R\,\rightarrow\,\infty}&\longrightarrow& -\Big(1-\frac{2M_{\scriptscriptstyle{\infty}}G}{R}\Big)\rd t^{2}+\Big(1+\frac{\,a-b\sqrt{1-h^{2}/b^{2}}\,}{R}\Big)\rd R^{2}+R^{2}\rd\Omega^{2}\,.
\ea
\label{largeRmetric}
\ee 
Hence, the rotation curve can be non-Keplerian only over a finite range. The second, with \eqref{explicitR}, gives the radius of a photon sphere (if positive). The last  is the  conserved global charge of   the time translational symmetry,  computable straightforwardly   following the prescription of  \cite{Park:2015bza} which generalized the Wald formalism~\cite{Wald:1993nt,Iyer:1994ys,Iyer:1995kg} to DFT and contains the $\ODD$ covariant extensions of both Noether potential and  boundary two-form (\textit{c.f.~}\cite{Blair:2015eba}). 
Appendix~\ref{APPCHARGE} contains  details. \\

\item \textbf{ Rotation curves around the spherically symmetric DFT-vacuum}

Various choices of the     parameters, $a,b,h$, are informative. 
\begin{itemize}
\item If we set ${h=0}$, our solution reduces to that of  F-JNW~\cite{Fisher:1948yn,Janis:1968zz}, and  further with  ${b=0}$  to  the Schwarzschild metric,
\be\textstyle{
\ba{ll}
\rd s^{2}=-(1-{a}/{R})\rd t^{2}+\frac{\rd R^{2}}{\,1-{a}/{R}\,}+R^{2}\rd\Omega^{2}\,,
\quad&\quad
\Vo=\sqrt{\frac{a}{2R}}\,.
\ea}
\label{Schwarzschild0}
\ee
~\\

\item If ${a=h=0}$, we reproduce the renowned orbital velocity formula proposed by Hernquist~\cite{Hernquist:1990be},
\be\textstyle{
\ba{lll}
\rd s^{2}=\frac{-\rd t^{2}+\rd R^{2}}{1+b/R}+R^{2}\rd\Omega^{2}\,,
\quad&\quad e^{2\phi}{=\frac{1}{1+b/R}}\,,\quad&\quad\Vo=\sqrt{\frac{bR}{2(R+b)^{2}}}\,.
\ea}
\ee
Remarkably, the orbital velocity is not monotonic; it assumes its maximum  value,  \,about  $35\%$ of the speed of light,\,   at $\textstyle{R=b}$,
\be
\textstyle{{\max[\Vo]=}\frac{1}{2\sqrt{2}}\simeq0.35}\,.
\ee 
~\\

\item If  ${h=0}$ and ${a=b}$,  we obtain with $\alpha=\frac{1}{\sqrt{2}}|a|\geq 0$,  
\be
\ba{ll}
\rd s^{2}=-\left(\frac{\sqrt{R^{2}+\alpha^{2}}-\alpha}{
\sqrt{R^{2}+\alpha^{2}}+\alpha}\right)^{\sqrt{2}}\rd t^{2}+\frac{R^{2}}{R^{2}+\alpha^{2}}\rd R^{2}+R^{2}\rd\Omega^{2}\,,\quad&\quad
e^{2\phi}=\left(\frac{\sqrt{R^{2}+\alpha^{2}}-\alpha}{
\sqrt{R^{2}+\alpha^{2}}+\alpha}\right)^{\frac{1}{\sqrt{2}}}\,,
\ea
\ee
and
\be
\Vo=\left(\frac{2\alpha^{2}}{R^{2}+\alpha^{2}}\right)^{\frac{1}{4}}\left(\frac{\sqrt{R^{2}+\alpha^{2}}-\alpha}{
R}\right)^{\sqrt{2}}\,.
\ee
The orbital velocity is maximal  at 
$\textstyle{R=(4+2\sqrt{6})^{\frac{1}{2}}\alpha}$ as
\be\textstyle{
\max\left[\Vo\right]=\left(\frac{2}{5+2\sqrt{6}\,}\right)^{\!\frac{1}{4}}\!\left(\frac{\sqrt{5+2\sqrt{6}\,}-1}{\sqrt{4+2\sqrt{6}\,}}\right)^{\!\sqrt{2}}\simeq\,\, 0.42\,.}
\ee
~\\

\item 
A yet more interesting limit  is the case of $a/b\rightarrow 0^{+}$ with    nontrivial $H$-flux, ${h\neq0}$. Especially when  ${a=0}$, we have
\be
{
\ba{ll}
\multicolumn{2}{l}{\rd s^{2}=e^{2\phi}\big(-\rd t^{2}+\frac{R^{2}}{\,R^{2}-h^{2}/4\,}\rd R^{2}\big)+R^{2}\rd\Omega^{2}\,,}\\
{e^{2\phi}=\frac{R_{h}^{2}}{R_{h}^{2}-1/2\,+\,\tan\upsilon\sqrt{R_{h}^{2}-1/4\,}}\,,}\quad&\quad
B_{\scriptscriptstyle{(2)}}=h\cos\vartheta\,\rd t\wedge\rd\varphi\,,\\
\multicolumn{2}{c}{\!\Vo\!={\frac{R_{h}}{{\left|R_{h}^{2}-{1}/{2}\,+\,\tan\upsilon\sqrt{R_{h}^{2}-{1}/{4\,}}\right|}}\!\!\left[{\frac{1}{2}\!\tan\upsilon\!\left(\frac{\,R_{h}^{2}-{1}/{2}\,}{\sqrt{R_{h}^{2}-{1}/{4\,}}\,}\right)\!-\frac{1}{2}}\right]^{\frac{1}{2}}},}
\ea}
\label{a0Vorbit}
\ee
where we set two dimensionless shorthand  variables,
\be
\textstyle{
\ba{ll}
R_{h}:={R}/{\left|h\right|}\,\geq\,1/2\,,\quad&\quad\tan\upsilon:=b\sqrt{h^{-2}-b^{-2}}\,.
\ea}
\ee
By tuning the  variable as ${\upsilon\rightarrow  0^{+}}$  (${h/b\rightarrow 1^{-}}$), it is possible to make the maximal velocity,  $\max[\Vo]$,    arbitrarily  small. Hence it may be   comparable to  observations;  it may simulate the galaxy rotation curve,  see Figure~\ref{FIGgrcMAIN}.\\

\begin{figure}[H]
\centering\includegraphics[width=90mm]{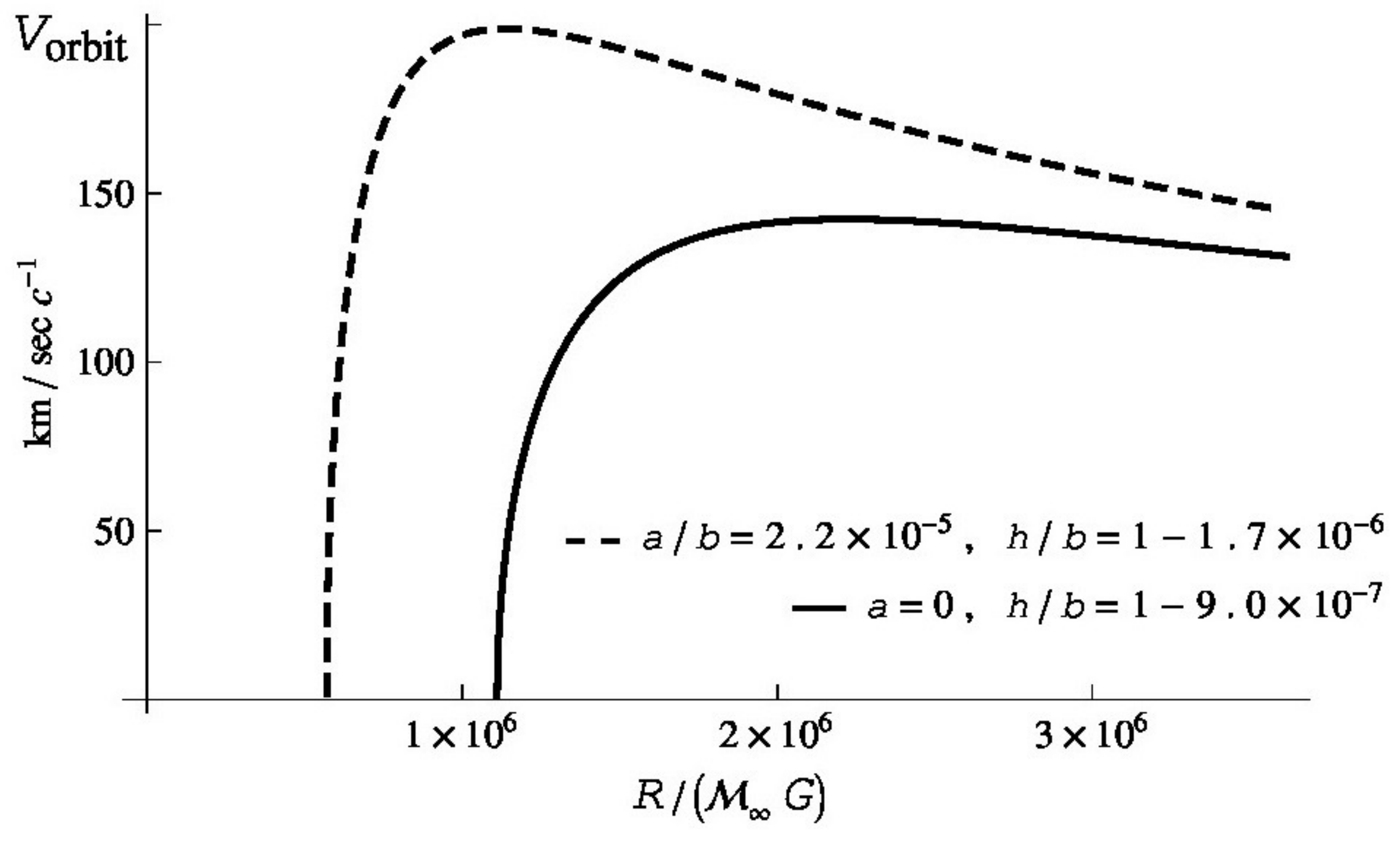}
\caption{\red{\bf{Rotation curves}} (dimensionless, nonexhaustive). 
The  curves with ${a/b\sim 0^{+}}$ and $\,{h/b\sim 1^{-}}$  feature a maximum of the orbital  velocity after   a fairly rapid  rise. It is   roughly  about $ 150\,\mbox{km/s}\,c^{-1}$ which is comparable to observations~\cite{Rubin:1980zd}.  Further,  if we let $R$ and  $M_{\scriptscriptstyle{\infty}}$ assume  the radius   and 
the   mass of the visible matter  in the Milky Way,  \textit{i.e.~}approximately   $15\,\mbox{kpc}$ and  ${2\times10^{11} M_{\odot}}$  respectively,    we have as an order of magnitude,    
$R/(M_{\scriptscriptstyle{\infty}}G)\simeq 1.5\times 10^{6}$. This number fits  our   scale of the  horizontal axis  above,  and  is  thousand times  smaller compared with  the Earth,  $R_{\oplus}/(M_{\oplus}G)\simeq 1.4\times 10^{9}$, \textit{c.f.~}`Cosmic Uroboros'.  For  small enough $R/(M_{\scriptscriptstyle{\infty}}G)$, the gravity becomes \textit{repulsive} and $\Vo$ is not defined (or pure imaginary).}
\label{FIGgrcMAIN}
\end{figure}

\end{itemize}

\end{itemize}

\newpage


\section{Discussion\label{SECCOM}} 
In this work of    theoretical interest,       we have aspired     to  assume DFT as  the stringy extension of  General Relativity. From the fundamental symmetry principle, such as  $\ODD$ covariance and doubled diffeomorphisms, we have  unambiguously determined the action for a point particle coupled to the closed string massless  sector. We have  showed that the  particle follows the geodesic set in   not  the Einstein  but   the  string frame.  We have analyzed the circular geodesic motion around the most general, spherically symmetric, asymptotically flat, static  ${D=4}$ DFT-vacuum.   Crucially, the resulting  rotation curve features  generically    a maximum and thus non-Keplerian over a  finite range (short-distance), while becoming  asymptotically Keplerian at infinity (long-distance),   all measured in terms of the dimensionless radial variable, $R/(M_{\scriptscriptstyle{\infty}}G)$,  which is  normalized by the mass in  natural units. Furthermore,  the gravitational force can be  even   repulsive quite close to the origin  (far-short-distance) [see (\ref{a0Vorbit}) and Figure~\ref{FIGgrcMAIN}].  By tuning the  three  free parameters of the spherically symmetric DFT-vacuum, such as  ${a/b\sim 0^{+}}$ and $\,{h/b\sim 1^{-}}$,  we have attempted   to  simulate   quantitatively, fitting  order of magnitude the scales of  both vertical and horizontal axes,  the   
 flat or slowly rising galaxy rotation curves  observed   for finite regions   outside  the  visible matter [Figure~\ref{FIGgrcMAIN}].        \\

\noindent While the proper radius, $R$, is the dimensionful physical radius, the normalized   radius,  $R/(M_{\scriptscriptstyle{\infty}}G)$, is the mathematically  natural dimensionless  variable which  essentially probes the theoretical nature of the gravitational force, \textit{not exclusively}, in  Double Field Theory. Intriguingly,   the normalized dimensionless  radius   is  thousand  times smaller for the Milky Way compared to the Earth  at  each surface (of  the visible matter):  $1.5\times 10^{6}$ versus $1.4\times 10^{9}$.  Note also $1 \mbox{AU}/(M_{\odot}G) \simeq 1.0\times 10^{8}$ for the solar system. 
\bgroup
\small
\def\arraystretch{1.1}%
\begin{center}
\begin{tabular}{|c|c|c|c|c|c|c|c|c|}
\hline
\!\!\!\!\!\!\!\!\!\!
\begin{tabular}{c}
\vspace{-5pt}
\includegraphics[width=19mm]{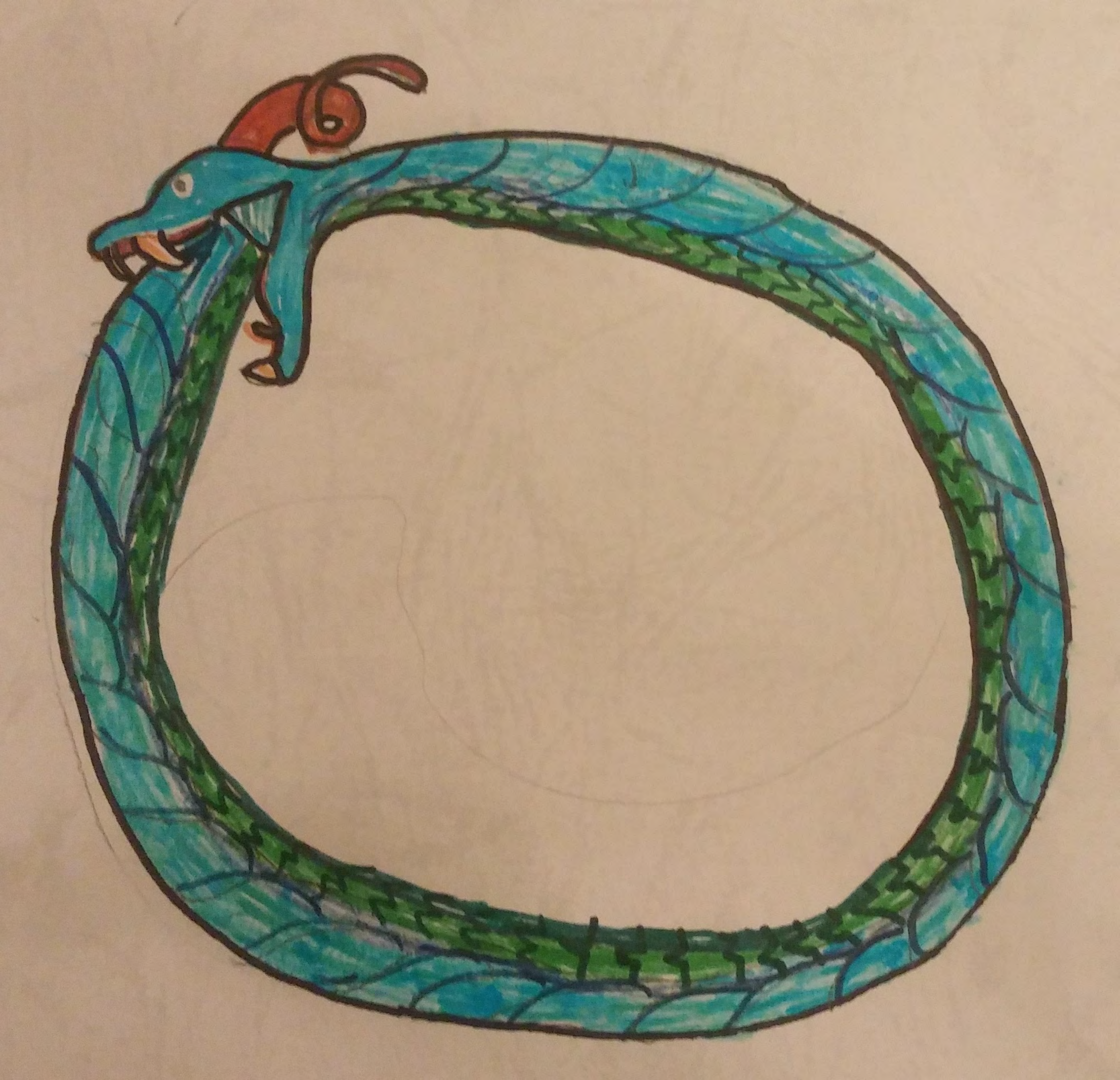}
\end{tabular}
\!\!\!\!\!\!\!\!\!
&
\!\!\!\!\begin{tabular}{c}
Electron\\ $(R\,{\simeq 0})$\end{tabular}
\!\!\!\!\!\!&
\!Proton\!&
\!\!\!\!\begin{tabular}{c}Hydrogen\\Atom\end{tabular}
\!\!\!\!&
\!\!Billiard Ball\!\!&
\!Earth\!&
\!\!\!\!\!\begin{tabular}{c}
Solar System\\ $(1{\rm{AU}}/M_{\odot}G)$\end{tabular}
\!\!\!\!\!\!&
\!\!\!\!\begin{tabular}{c}Milky Way\\(visible)\end{tabular}
\!\!\!\!\!&
\!\!\!\!\!\begin{tabular}{c}
Universe\\ $(M_{\scriptscriptstyle{\infty}}{\propto R^{3}})$\end{tabular}
\!\!\!\!\!\\
\hline
$\!\!\blue{R/(M_{\scriptscriptstyle{\infty}}G)}\!\!$& 
\!\!$0^{+}$\!\! &
\!\! $7.1{\times10^{38}}$\!\! &
\!\!$2.0{\times 10^{43}}$\!\! & 
\!\!$2.4{\times 10^{26}}$\!\! &
\!\! $1.4{\times 10^{9}}$\!\! &
\!\!$1.0{\times 10^{8}}$\!\! & 
\!\!$1.5{\times 10^{6}}$\!\! & 
\!\!$0^{+}$\!\!\\
\hline
\end{tabular}
~\\
~\\
\textbf{`Uroboros' spectrum  of the dimensionless radial variable normalized by mass in natural units. \\
The orbital speed    is also   dimensionless, and   depends  on   the single  variable, $R/(M_{\scriptscriptstyle{\infty}}G)$.   }
\label{Table}
\end{center}
\egroup

Generically, if the mass density  is  constant,    the  dimensionless radial variable, $R/(M_{\scriptscriptstyle{\infty}}G)$, becomes smaller as the physical radius, $R$, grows.  This  seems to   imply  that,  the  observations of stars and galaxies  far away, or the dark matter and the dark energy  problems, are actually revealing   the short-distance nature of gravity, as they are essentially based on  small $R/(M_{\scriptscriptstyle{\infty}}G)$ observations (long distance divided by  far heavier mass).   Perhaps, the repulsive gravitational force at very short-distance,  $R/(M_{\scriptscriptstyle{\infty}}G)\rightarrow 0^{+}$, might be responsible for the inflation or  the accelerating expansion of the Universe. We believe this speculation of solving  the dark matter/energy problems by modifying short gravity   deserves further explorations, even  not necessarily restricted to  the framework of Double Field Theory. \\

\noindent  From the coupling of the closed string massless sector to the Standard Model~(\ref{minimalcouplingF}), (\ref{minimalcouplingB}),  the $B$-field (or the axion) couples to the fermions only: it  does not interact with any gauge bosons, and hence  transparent, or dark,  to electromagnetic radiation.  In contrast, the scalar dilaton, $\phi$,   couples to the bosons only but not to any  fermion, $\chi$ in  (\ref{minimalcouplingF}).   
 As the scalar dilaton, $\phi$,  and the $B$-field   are ``massless'', they tend to spread  over larger space and get diluted, but not completely, as our asymptotically flat  solution is anyhow non-Keplerian   up to finite range,    $R/(M_{\scriptscriptstyle{\infty}}G)<<\infty$. It is worth while   to note that, in the  string frame the scalar field, $\Phi:=e^{-\phi}$, acquires an effective mass given by the scalar curvature  as  $4\partial_{\mu}\Phi\partial^{\mu}\Phi+R\Phi^{2}$ in the Lagrangian. While   DFT modifies the law of Einstein gravity, from the conventional GR point of view, the scalar  dilaton and the $B$-field may well be then regarded as   extra `dark matter' (\textit{c.f.~}axion~\cite{Kim:1979if,Shifman:1979if,Dine:1981rt,Zhitnitsky:1980tq}), or `dark gravity' (as part of stringy gravity). This identification appears    consistent with  the `bullet cluster' observation~\cite{Clowe:2006eq} which often rules out  theories of modified gravity.   Furthermore, with the identification of     the asymptotic mass, $M_{\scriptscriptstyle{\infty}}$ defined in \eqref{observables},   as the (baryonic)  mass of the visible matter,   it is worth while to note  that even if $M_{\scriptscriptstyle{\infty}}$ vanishes, there exists  a class of  nontrivial DFT-vacuum solutions. This might also explain the observed gravitational lensing without visible matter. \\

\noindent  Certainly, the   phenomenological validity of DFT, as an alternative to GR,  is still questionable, requires and  deserves  further thorough  verifications.    Compared to other theories of modified gravity,  DFT is  singled out as the string theory extension of Einstein gravity guided entirely by the symmetry principle: the fundamental symmetries of $\ODD$ T-duality, doubled diffeomorphisms and twofold local Lorentz symmetries rigidly fix the theory,  including   the couplings  to the Standard Model and to a point-like particle.   Thus,   while testing   DFT    against more high-precision data of  observations in future,  it should be taken into account   that  \textit{i)} a relativistic point particle follows the geodesic motion not in the Einstein frame but in the string frame, and \textit{ii)} the scalar dilaton and the $B$-field are transparent or `dark' to the Standard Model fermions and the gauge bosons respectively.  Deeper understanding of the three parameters, perhaps as the intrinsic properties of matter or an elementary particle, would be desirable.  For this,  once again,  the minimal coupling between DFT and the Standard Model~(\ref{minimalcouplingF}), (\ref{minimalcouplingB}), could be a starting point. 

\section*{Acknowledgements}
We wish to thank David Berman, Nakwoo Kim, Wontae Kim,  Soo-Jong Rey, Yuho Sakatani, Chang Sub Shin,  Hyun Seok Yang  and Sang-Heon Yi for helpful discussions.  This work  was  supported by  the National Research Foundation of Korea   through  the Grants   
2015K1A3A1A21000302  and  2016R1D1A1B01015196.\\

\appendix
\begin{center}
\huge\textbf{APPENDIX}
\end{center}

\section{More rotation curves around  various spherically symmetric DFT-vacua}
Here, for various choices of the  free parameters, $\{a,b,h\}$~\eqref{THESOL0}, we depict the  corresponding rotation curve  as a plot of the two  \textit{dimensionless} quantities, namely   the orbital velocity, $\Vo$,  versus the  scaled  proper radius,   $R/(M_{\scriptscriptstyle{\infty}}G)$, where  the `asymptotic' mass, $M_{\scriptscriptstyle{\infty}}$, is defined in \eqref{observables}.

\begin{itemize}
\item If $b=h=0$ and $a=2M_{\scriptscriptstyle{\infty}}G>0$, we recover the  Schwarzschild metric, as  in \eqref{Schwarzschild0},    
\be{
\ba{lll}
\rd s^{2}=-\left(1-\frac{2M_{\scriptscriptstyle{\infty}}G}{R}\right)\rd t^{2}+\left(1-\frac{2M_{\scriptscriptstyle{\infty}}G}{R}\right)^{-1}\rd R^{2}+R^{2}\rd\Omega^{2}\,,
~~~&~~~\phi=0\,,~~~&~~~ B_{\mu\nu}=0\,.
\ea}
\label{Schwarzschild}
\ee
The corresponding Keplerian rotation curve is depicted in  Figure~\ref{FIGSchwarzschild}.\\

\begin{figure}[H]
\centering\includegraphics[width=85mm]{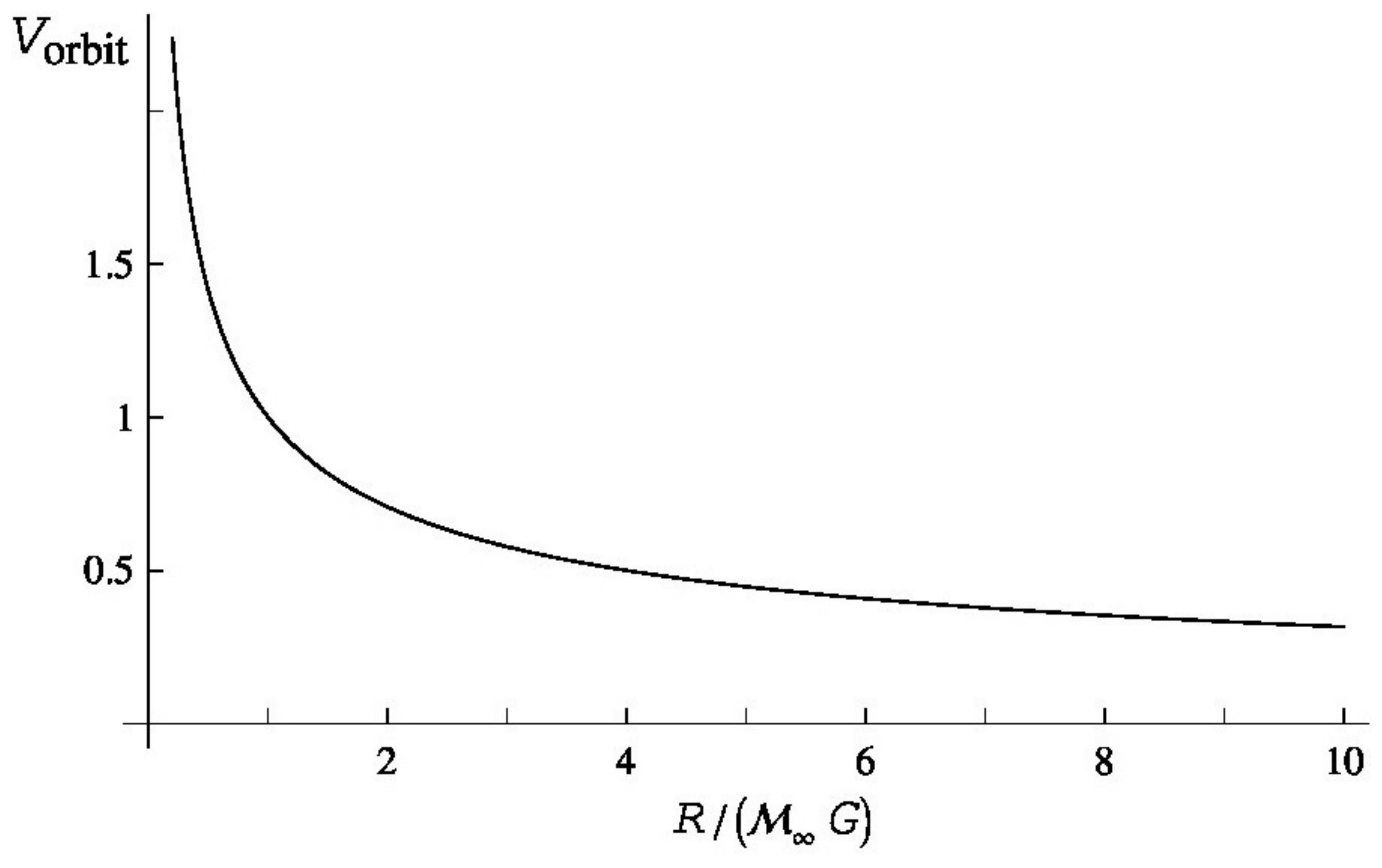}
\caption{{\bf{\red{Keplerian rotation curve of the Schwarzschild geometry, $\Vo=\sqrt{\frac{M_{\scriptscriptstyle{\infty}}G}{R}}$~:}}} {{$\blue{b=h=0}$}}}
\label{FIGSchwarzschild}
\end{figure}
~\\

If $b=h=0$ and $a=2M_{\scriptscriptstyle{\infty}}G<0$,   we have  
\be\textstyle{
\ba{lll}
\rd s^{2}=-\left(1+\frac{2M_{\scriptscriptstyle{\infty}}G}{r}\right)^{-1}\rd t^{2}+\left(1+\frac{2M_{\scriptscriptstyle{\infty}}G}{r}\right)\rd r^{2}+(r+2M_{\scriptscriptstyle{\infty}}G)^{2}\rd\Omega^{2}\,,
~~&~~\phi=0\,,~~&~~ B_{\mu\nu}=0\,,
\ea}
\ee
which, after the radial coordinate redefinition, $r\rightarrow R-2M_{\scriptscriptstyle{\infty}}G$,  reduces to the Schwarzschild metric~(\ref{Schwarzschild}), yet with the negative mass.\\

\item If $a=h=0$, regardless of the sign of $b$ (up to possible  radial coordinate shift),  we get  
\be\textstyle{
\ba{lll}
\rd s^{2}=-\frac{R}{R+b}\rd t^{2}+\frac{R}{R+b}\rd R^{2}+R^{2}\rd\Omega^{2}\,,\quad
&\quad e^{2\phi}=\frac{R}{R+b}\,,\quad&\quad B_{\mu\nu}=0\,.
\ea}
\label{zerob}
\ee
The corresponding  orbital velocity coincides with that of the  Hernquist model~\cite{Hernquist:1990be} up to an overall constant factor [see Figure~\ref{FIGa0b1h0}], 
\be{
\Vo=\sqrt{\frac{bR}{2(R+b)^{2}}}\,,}
\label{Voahz}
\ee
which is valid for positive $b$. Otherwise the gravity is repulsive.  The orbital velocity would have been trivial if we had computed it in the Einstein frame where the temporal component of the Einstein frame metric is constant,  `$g^{\scriptscriptstyle{\rm{E}}}_{tt}=1$'.\\

\begin{figure}[H]
\centering\includegraphics[width=85mm]{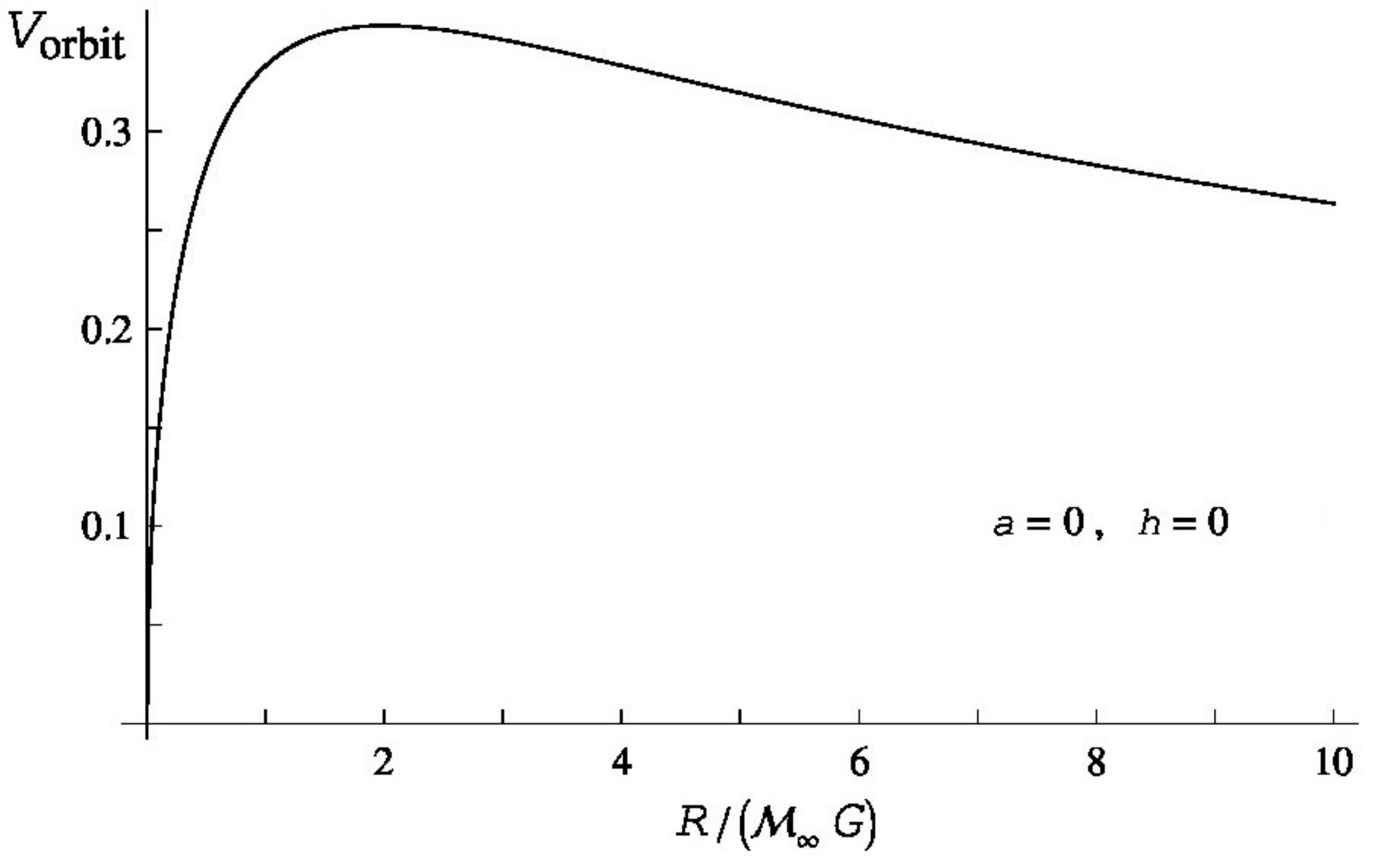}
\caption{{\bf{\red{Rotation curve of $~\Vo=\sqrt{\frac{bR}{2(R+b)^{2}}}$~:}}} \blue{Hernquist Model~\cite{Hernquist:1990be}, {{$\blue{a=h=0}\,$.}}} The  orbital velocity  assumes its maximum value, $\max\left[\Vo\right]=\frac{1}{\,2\sqrt{2}\,}\simeq 0.353553\,$ (about $35$\% of the speed of light)  at $R=b=2M_{\scriptscriptstyle{\infty}}G$. }
\label{FIGa0b1h0}
\end{figure}

\item If $h=0$, we recover the  F-JNW solution~\cite{Fisher:1948yn,Janis:1968zz} which is the   most general static spherical solution of the Einstein gravity coupled to the scalar dilaton,
\be
\textstyle{\ba{ll}
\multicolumn{2}{l}{\rd s^{2}=-\left(1-\frac{\sqrt{a^{2}+b^{2}}}{r}\right)^{\frac{a+b}{\sqrt{a^{2}+b^{2}}}}\rd t^{2}+
\left(1-\frac{\sqrt{a^{2}+b^{2}}}{r}\right)^{\frac{-a+b}{\sqrt{a^{2}+b^{2}}}}\left[
\rd r^{2}+
r\left(r-\sqrt{a^{2}+b^{2}}\right)\rd\Omega^{2}\right]\,,}\\
e^{2\phi}=
\left(1-\frac{\sqrt{a^{2}+b^{2}}}{r}\right)^{\frac{b}{\sqrt{a^{2}+b^{2}}}}\,,\qquad&\qquad B_{\mu\nu}=0\,.
\ea}
\ee
In this case of $h=0$, the T-duality over the temporal direction, $t\leftrightarrow\tilde{t}$ (\textit{c.f.~}\cite{Bergshoeff:2011se,Malek:2013sp}) preserves the form of the F-JNW solution given in the string frame,   and it results in exchanging the two  parameters, $(a,b)\leftrightarrow(-b,-a)$.

The solution can be also rewritten, after the shift,  $r\rightarrow\textstyle{r+\frac{b}{a+b}\sqrt{a^{2}+b^{2}}}$,  as
\[
\rd s^{2}=-\left(\frac{r-\alpha}{r+\beta}\right)^{\frac{a+b}{\sqrt{a^{2}+b^{2}}}}\rd t^{2}+\left(\frac{r-\alpha}{r+\beta}\right)^{\frac{-a+b}{\sqrt{a^{2}+b^{2}}}}\left[{\small{\rd r^{2}+
\left(r-\alpha\right)
\left(r+\beta\right)\rd\Omega^{2}}}\right]\,,
\]
\be
\ba{ll}
e^{2\phi}=
\left(\frac{r-\alpha}{r+\beta}\right)^{\frac{b}{\sqrt{a^{2}+b^{2}}}}\,,\quad&\quad B_{\mu\nu}=0\,.
\ea
\ee
which manifestly interpolates (\ref{Schwarzschild}) and (\ref{zerob}), in a unifying manner.\\

Figures~\ref{FIGa1b2h0} and \ref{FIGa2b1h0} show the orbital velocities for the choices of  the parameters, $a/b=0.5$ and  $a/b=2$,   commonly  with $h=0$.\\
~\\

\begin{figure}[H]
\centering
\begin{minipage}{.5\textwidth}
  \centering
\includegraphics[width=80mm]{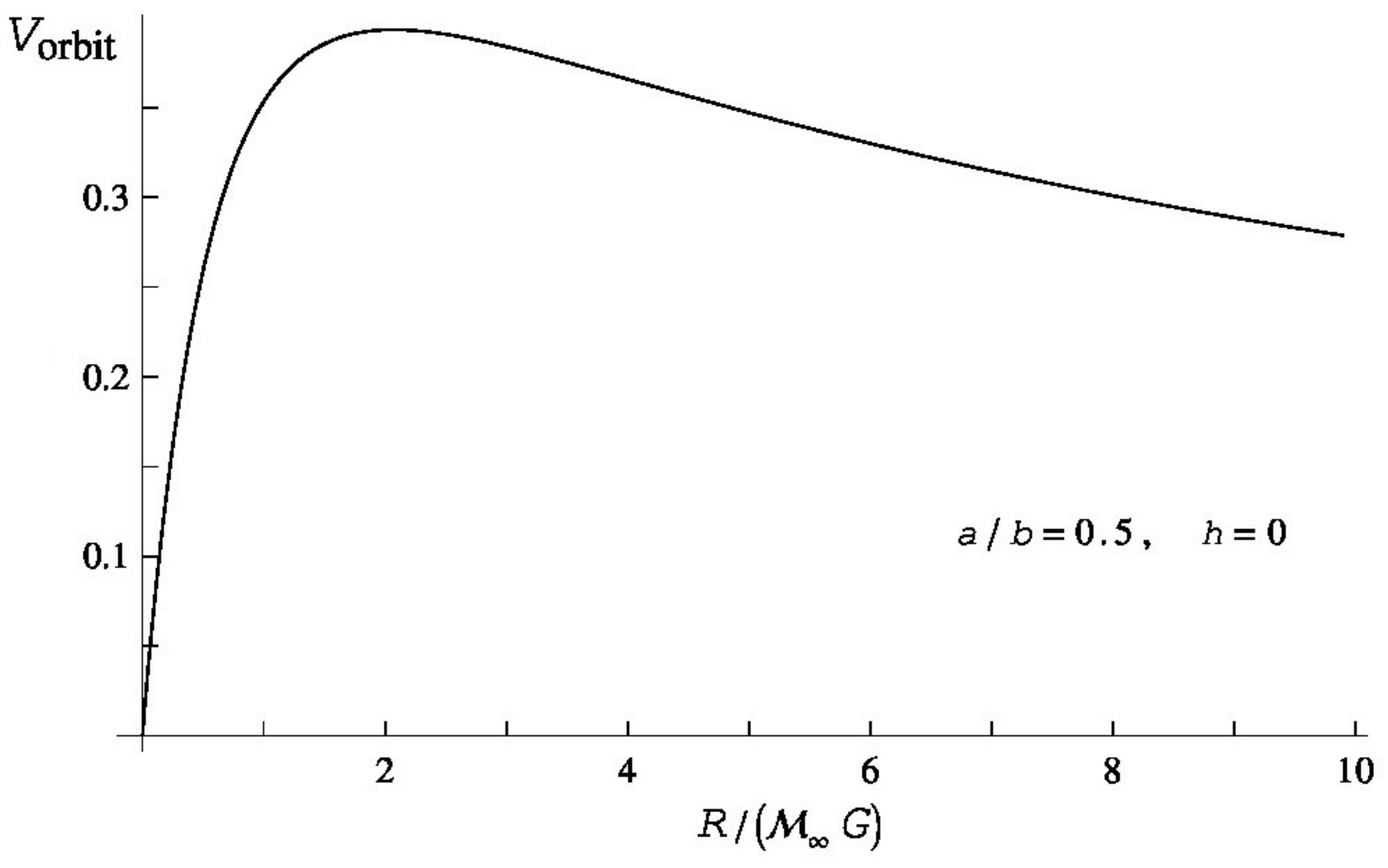}
\caption{{\bf{\red{Rotation curve:}}} {{$\blue{a/b=0.5, \,~h=0}$}}}
\label{FIGa1b2h0}
\end{minipage}%
\begin{minipage}{.5\textwidth}
\centering
\includegraphics[width=80mm]{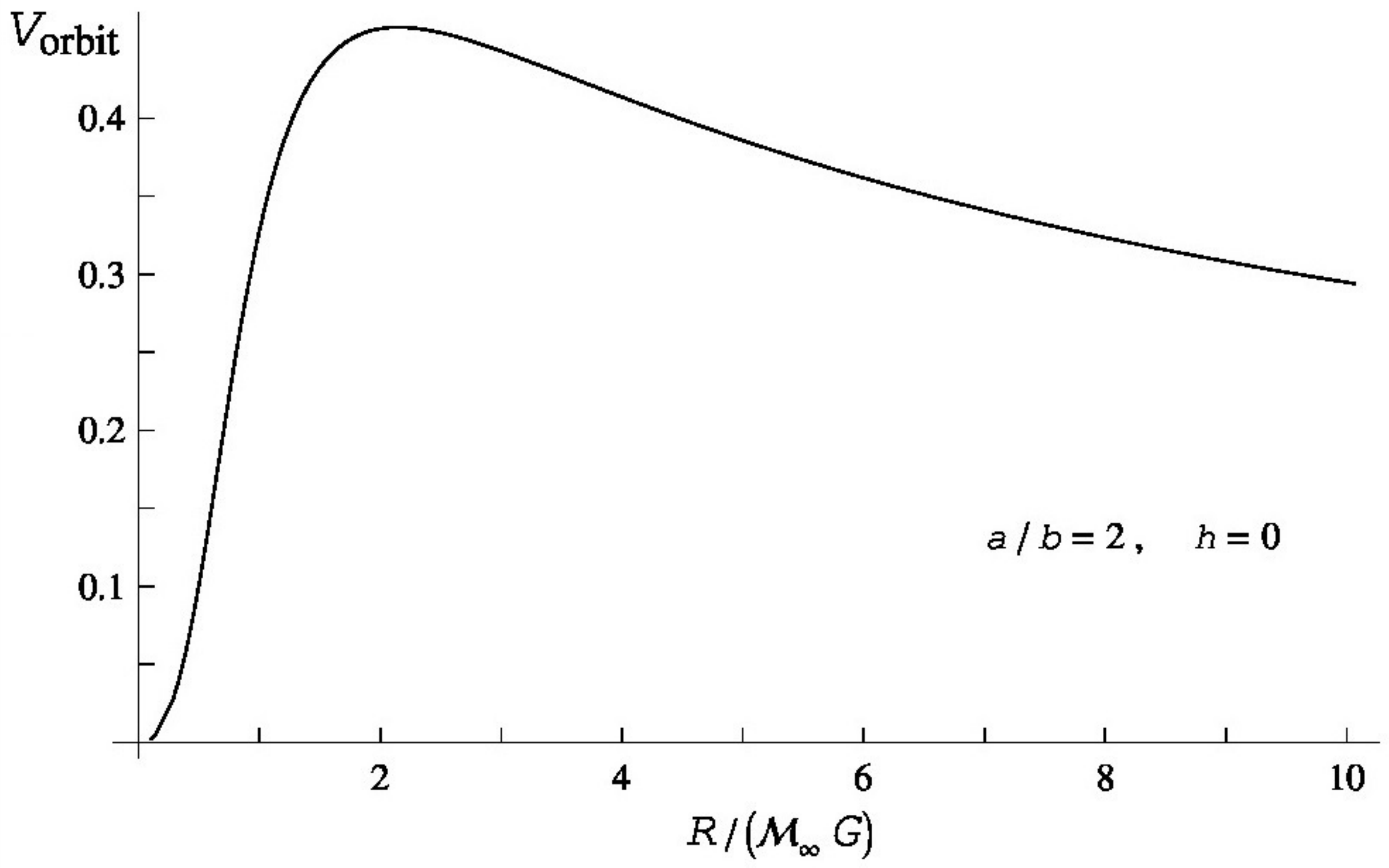}
\caption{{\bf{\red{Rotation curve:}}} {{$\blue{a/b=2,  ~ \,h=0}$}}}
\label{FIGa2b1h0}
\end{minipage}%

\end{figure}


\item If $h=0$ and ${a=b}$, with the proper radius, $R\equiv\sqrt{r^{2}-\alpha^{2}}$,  and a positive number, $\alpha\equiv\frac{1}{\sqrt{2}}|a|>0$,  the above solution reduces to 
\be
{\ba{lll}
\rd s^{2}=-\left(\frac{\sqrt{R^{2}+\alpha^{2}}-\alpha}{
\sqrt{R^{2}+\alpha^{2}}+\alpha}\right)^{\sqrt{2}}\rd t^{2}+\frac{R^{2}}{R^{2}+\alpha^{2}}\rd R^{2}+R^{2}\rd\Omega^{2}\,,\quad&
e^{2\phi}=\left(\frac{\sqrt{R^{2}+\alpha^{2}}-\alpha}{
\sqrt{R^{2}+\alpha^{2}}+\alpha}\right)^{\frac{1}{\sqrt{2}}}\,,\quad& B_{\mu\nu}=0\,.
\ea}
\ee
The orbital velocity is [see Figure~\ref{FIGa1b1h0}],
\be
{
\Vo=\left(\frac{2\alpha^{2}}{R^{2}+\alpha^{2}}\right)^{\frac{1}{4}}\left(\frac{\sqrt{R^{2}+\alpha^{2}}-\alpha}{
R}\right)^{\sqrt{2}}\,,}
\ee
which assumes its maximum value, about $42$\% of the speed of light,  at 
$R=(4+2\sqrt{6})^{\frac{1}{2}}\alpha\,$:
\be
\textstyle{
\max\left[\Vo\right]=\left(\frac{2}{5+2\sqrt{6}\,}\right)^{\frac{1}{4}}\left(\frac{\sqrt{5+2\sqrt{6}\,}-1}{\sqrt{4+2\sqrt{6}\,}}\right)^{\sqrt{2}}\simeq 0.420868\,.}
\ee
~\\

\begin{figure}[H]
\centering\includegraphics[width=85mm]{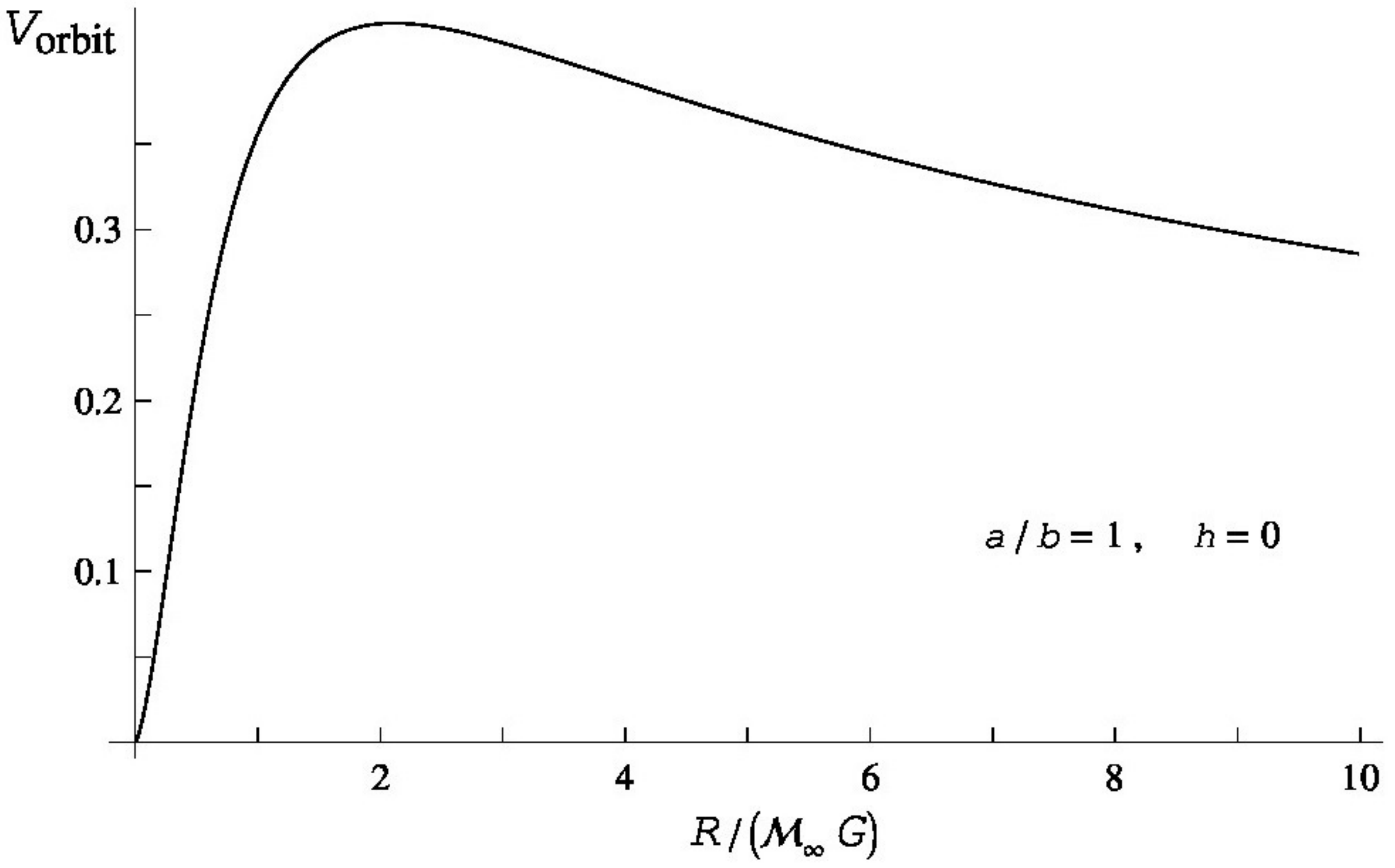}
\caption{{\bf{\red{Rotation curve:}}} {{$\blue{a/b=1,  ~ \,h=0}$}}}
\label{FIGa1b1h0}
\end{figure}
~\\

\item If ${a=0}$ with ${b^{2}\geq h^{2}}$,  up to some alternative radial coordinate shift,  we  get
\be
\ba{lll}
\rd s^{2}&=&e^{2\phi}\left(-\rd t^{2}+\rd r^{2}\right)+(r^{2}+\frac{1}{4}h^{2})\,\rd\Omega^{2}\,,\\
e^{2\phi}&=&\frac{4r^{2}+h^{2}}{4r^{2}\pm 4r\sqrt{b^{2}-h^{2}}\,-h^{2}}\,,\\
H_{\scriptscriptstyle{(3)}}&=&h\sin\vartheta\,\rd t\wedge\rd\vartheta\wedge\rd\varphi\,,
\ea
\label{zeroa}
\ee
where the free sign, $\pm$, coincides with that of $b$. This solution can be  rewritten in terms of the proper radius satisfying  $R^{2}= r^{2}+\frac{1}{4}h^{2}$, to take the form:
\be 
\ba{lll}
\rd s^{2}&=&e^{2\phi}\left(-\rd t^{2}+\frac{R^{2}}{R^{2}-{h^{2}/4}}\rd R^{2}\right)+R^{2}\rd\Omega^{2}\,,\\
e^{2\phi}&=&\frac{R^{2}}{R^{2}-{h^{2}/2}\,\pm\sqrt{b^{2}-h^{2}}\sqrt{R^{2}-{h^{2}/4}}}\,,\\
H_{\scriptscriptstyle{(3)}}&=&h\sin\vartheta\,\rd t\wedge\rd\vartheta\wedge\rd\varphi\,.
\ea
\label{zeroa2}
\ee
Requiring the reality, we need to  constrain the range of the proper radius,  at least,  
\be
R\geq\half|h|\,.
\label{range}
\ee
The orbital velocity is  [see Figures~\ref{FIGa0b2h1} and \ref{FIGa0b1hl}],
\be
\textstyle{
\Vo=\frac{R}{\,\left|\,R^{2}-\frac{1}{2}h^{2}\pm\sqrt{b^{2}-h^{2}}\sqrt{R^{2}-\frac{1}{4}h^{2}}\,\right|\,}\sqrt{\pm\sqrt{\frac{b^{2}-h^{2}}{4R^{2}-h^{2}}\,}\left(\textstyle{R^{2}-\frac{1}{2}h^{2}}\right)-\textstyle{\frac{1}{2}h^{2}}}\,,}
\ee
which  reduces to (\ref{Voahz}) when $h=0$ for consistency.\,       
If $h\neq0$, it can be rewritten  as
\be
\textstyle{
\Vo={\frac{R_{h}}{{\left|\,R_{h}^{2}-\frac{1}{2}+\tan\upsilon\sqrt{R_{h}^{2}-\frac{1}{4}}\,\right|}}\left[{\frac{1}{2}\tan\upsilon\left(\frac{\,R_{h}^{2}-\frac{1}{2}\,}{\sqrt{R_{h}^{2}-\frac{1}{4}}}\right)-\frac{1}{2}}\right]^{{1/2}}}\,,}
\ee
for which we set dimensionless variables,
\be
\ba{lll}
R_{h}:={R}/{\left|h\right|}\,,\quad&\quad \cos\upsilon:=\left|h\right|/b\,,\quad&\quad\sin\upsilon:=\sqrt{1-h^{2}/b^{2}}\,\geq\, 0\,.
\ea
\ee
As  for Figures~\ref{FIGgrcMAIN} and \ref{FIGa0b1hl},  it is worth while to note the dimensionless unit, $ 100\,\mbox{km/s}\,c^{-1}\simeq 3.336\times 10^{-4}$.\\


\begin{figure}[H]
\centering
\begin{minipage}{.5\textwidth}
  \centering
\includegraphics[width=80mm]{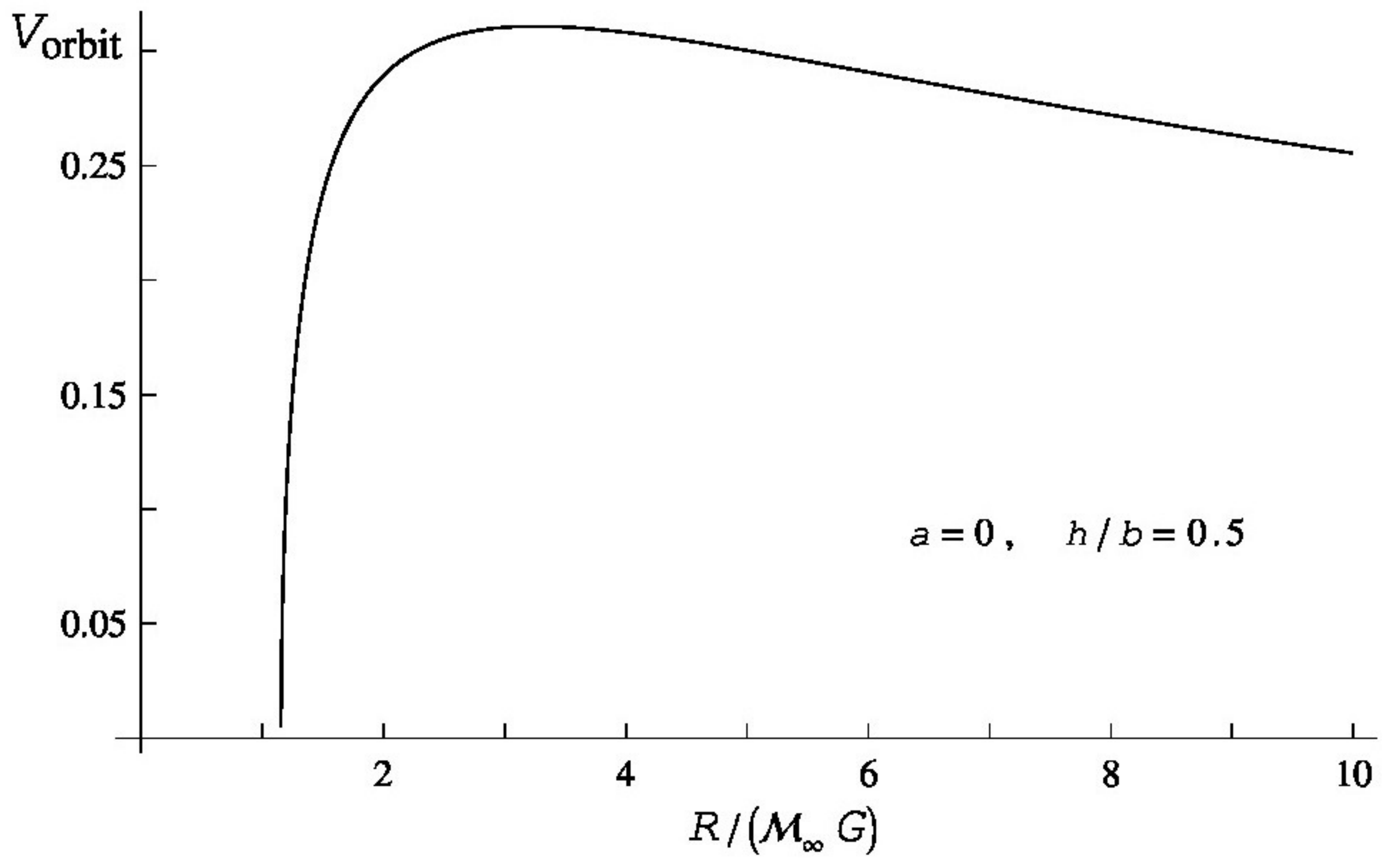}
\caption{{\bf{\red{Rotation curve:}}} {{$\blue{a=0,~ \,h/b=0.5}$}}}
\label{FIGa0b2h1}
\end{minipage}%
\begin{minipage}{.55\textwidth}
\centering
\includegraphics[width=80mm]{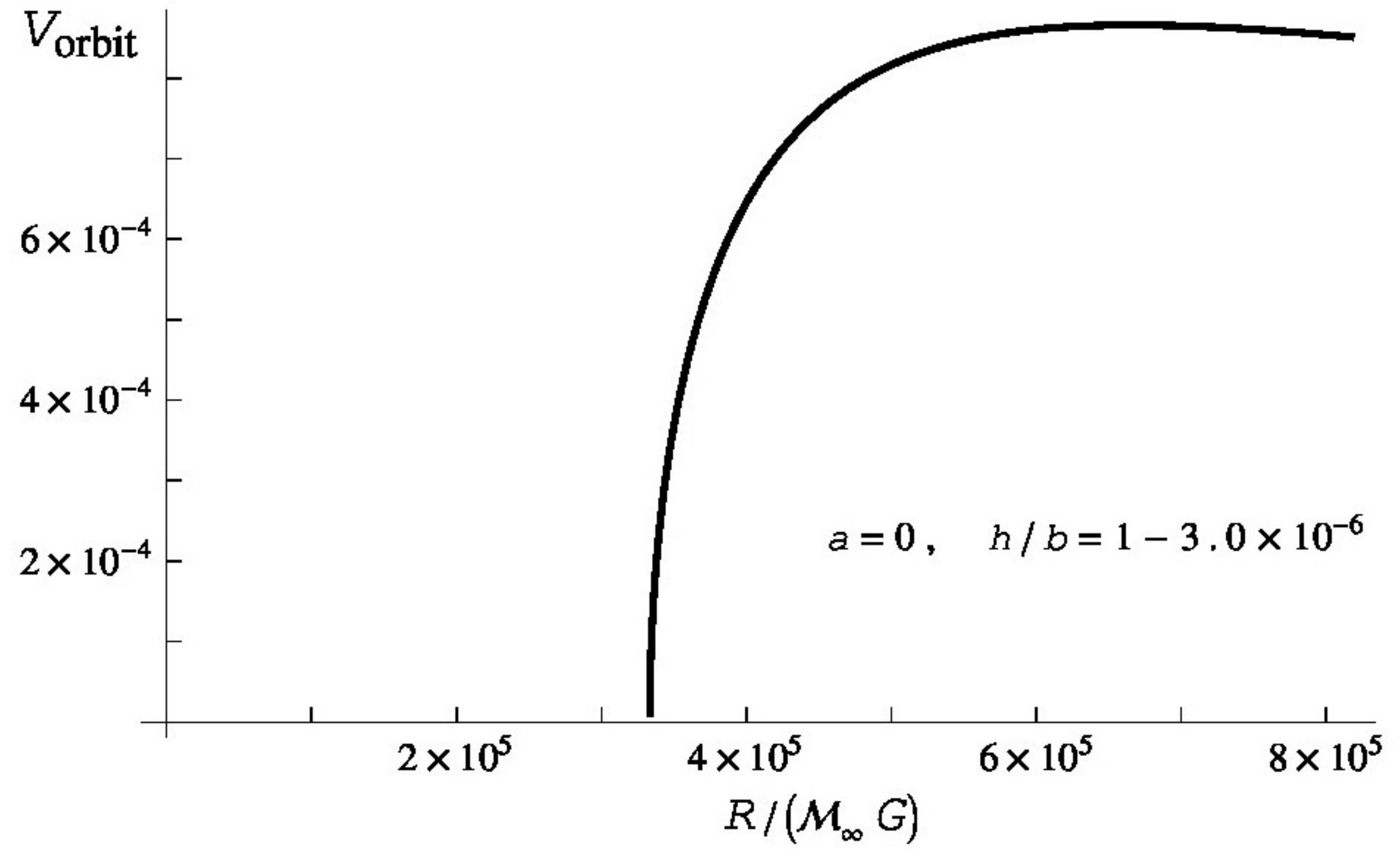}
\caption{{\bf{\red{Rotation curve:}}} {{$\blue{a=0,~ \,h/b=1-3.0{\times10^{-6}}}$}}}
\label{FIGa0b1hl}
\end{minipage}%
\end{figure}

\item In particular, if $a=0$ and $b^{2}=h^{2}$ saturated, the above solution~(\ref{zeroa2})  reduces to
\be
{\ba{lll}
\rd s^{2}&=&-\,\frac{R^{2}}{R^{2}-\frac{1}{2}h^{2}}\rd t^{2}+
\frac{R^{4}}{(R^{2}-\frac{1}{2}h^{2})(R^{2}-\frac{1}{4}h^{2})}\rd R^{2}+R^{2}\rd\Omega^{2}\,,\\
e^{2\phi}&=&\frac{R^{2}}{R^{2}-\frac{1}{2}h^{2}}\,,\\
H_{\scriptscriptstyle{(3)}}&=&h\sin\vartheta\,\rd t\wedge\rd\vartheta\wedge\rd\varphi\,.
\ea}
\ee
It is  no longer necessary to impose the constraint~(\ref{range}). Yet,  the gravity is now  repulsive and the orbital velocity becomes imaginary   making no physical  sense,
\be
{\Vo={\textstyle{\sqrt{-\frac{1}{2}}}}\times\frac{hR}{\left|R^{2}-\frac{1}{2}h^{2}\right|}\,.}
\ee

\item For a generic case with  all non-vanishing parameters, $a,b,h$, we may plot the corresponding  rotation curve numerically, based on the exact expressions of $R(r)$ and $\Vo(r)$, \eqref{explicitR} and \eqref{RvarphiV} respectively, see Figues\,\ref{FIGgrcMAIN} and \ref{FIGalb1h0}.\\

\begin{figure}[H]
\centering\includegraphics[width=85mm]{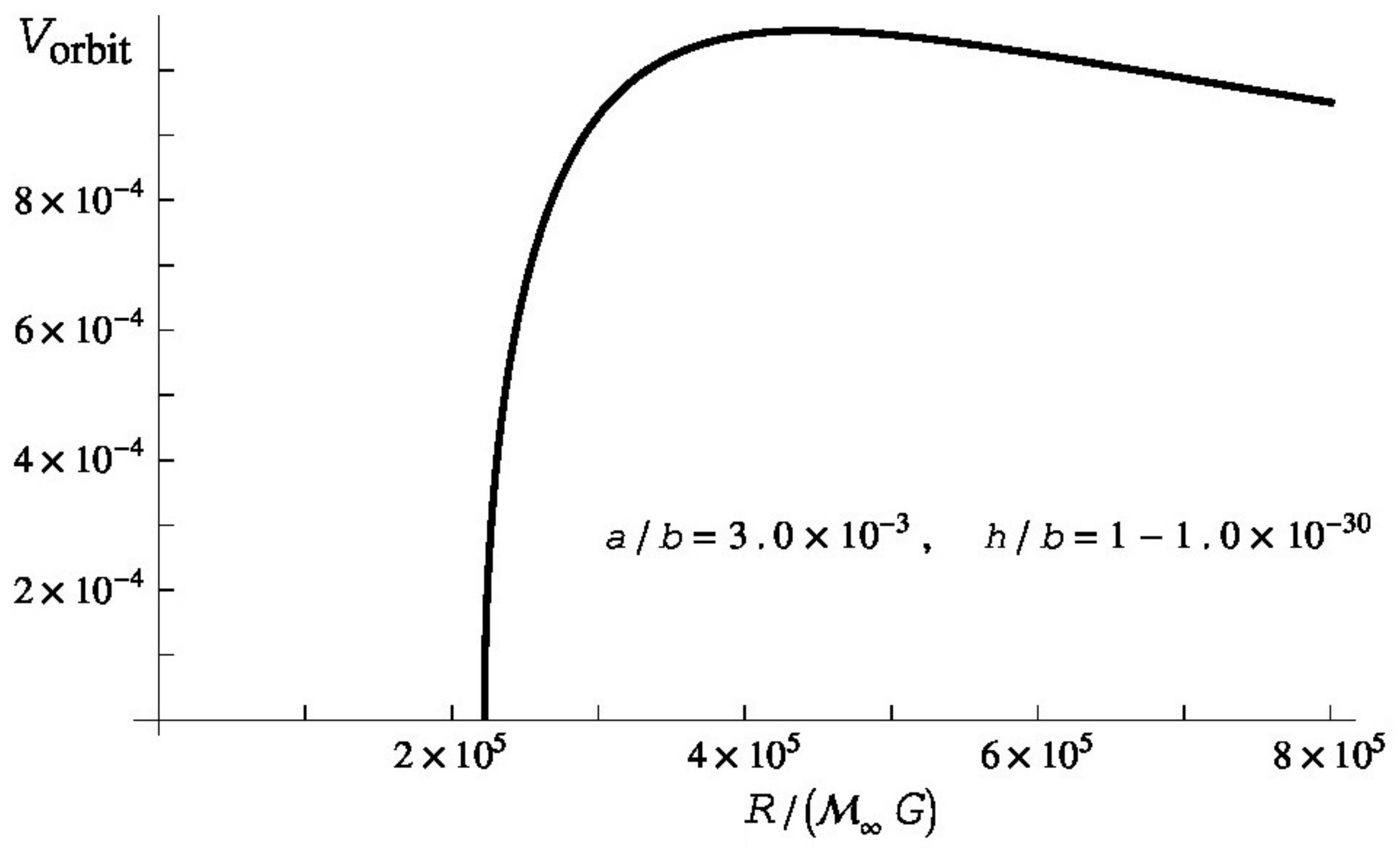}
\caption{{\bf{\red{Rotation curve:}}} {{$\blue{a/b=3.0\times 10^{-3},  ~ \,h/b=1-1.0\times 10^{-30}}$}}}
\label{FIGalb1h0}
\end{figure}

\end{itemize}
\section{Derivation of the   spherically symmetric  vacuum  solution to $D=4$ DFT\label{APPSOL}}

Without loss of generality, utilizing the radial diffeomorphisms, we assume the following  static, spherically symmetric ansatz for the string frame  metric,
\be
\rd s^{2}=e^{2\phi(r)}\big[-A(r)\rd t^{2} +A^{-1}(r)\rd r^{2}+A^{-1}(r)C(r)\rd\Omega^{2}\,\big]\,,
\label{metric}
\ee
where we put as  shorthand  notation, 
\be
\rd\Omega^{2}=\rd\vartheta^{2}+\sin^{2}\vartheta\,\rd\varphi^{2}\,.
\ee
It is worth while to note that our string frame metric ansatz takes  the  product  form of  the dilaton factor, $e^{2\phi}$, times the  Einstein frame metric.\\

\noindent If the spacetime  is  asymptotically `flat',   our metric  asnatz~(\ref{metric}) should meet  boundary conditions,
\be
\ba{lll}
\dis{\lim_{r\rightarrow\infty}A(r)=1\,,}\quad&\quad
\dis{\lim_{r\rightarrow\infty}r^{-2}C(r)=1\,,}\quad&\quad
\dis{\lim_{r\rightarrow\infty}\phi(r)=0\,,}
\ea
\label{BC}
\ee
and also from the asymptotic `smoothness',
\be
\ba{lll}
\dis{\lim_{r\rightarrow\infty}{A^{\prime}}(r)=\lim_{r\rightarrow\infty}{A^{\prime\prime}}(r)=0\,,}~~&~
\dis{\lim_{r\rightarrow\infty}r^{-1}{C^{\prime}}(r)=\lim_{r\rightarrow\infty}{C^{\prime\prime}}(r)=2\,,}~~&~
\dis{\lim_{r\rightarrow\infty}{\phi^{\prime}}(r)=\lim_{r\rightarrow\infty}{\phi^{\prime\prime}}(r)=0\,.}
\ea
\label{BC2}
\ee
We often write the $B$-field using  the form notation,
\be
B_{\scriptscriptstyle{(2)}}=\half B_{\mu\nu}\rd x^{\mu}\wedge\rd x^{\nu}=
B(r)\cos\vartheta\,\rd r\wedge\rd\varphi+ h\cos\vartheta\,\rd t\wedge\rd\varphi\,,
\ee
such that its field strength, or the $H$-flux, takes the most general spherically symmetric form,
\be
H_{\scriptscriptstyle{(3)}}=\textstyle{\frac{1}{3!}}H_{\lambda\mu\nu}\rd x^{\lambda}\wedge\rd x^{\mu}\wedge\rd x^{\nu}=
B(r)\sin\vartheta\,\rd r\wedge\rd\vartheta\wedge\rd\varphi+ h\sin\vartheta\,\rd t\wedge\rd\vartheta\wedge\rd\varphi\,,
\ee
which is closed for constant $h$. It is  spherically symmetric as it admits the   $\so(3)$ Killing vectors given by the usual  angular momentum differential operators, 
\be
\ba{lll}
\multicolumn{3}{c}{
\cL_{\xi_{a}}H_{\scriptscriptstyle{(3)}}=\rd\left(  \mathbf{i}_{\xi_{a}}H_{\scriptscriptstyle{(3)}}\right)
+\mathbf{i}_{\xi_{a}}\left(\rd H_{\scriptscriptstyle{(3)}}\right)=0\,,}\\
\xi_{1}=\sin\varphi\partial_{\vartheta}+\cot\vartheta\cos\varphi\partial_{\varphi}\,,\quad&\quad
\xi_{2}=-\cos\varphi\partial_{\vartheta}+
\cot\vartheta\sin\varphi\partial_{\varphi}\,,\quad&\quad
\xi_{3}=-\partial_{\varphi}\,,\\
\multicolumn{3}{c}{\big[\xi_{a}\,,\,\xi_{b}\big]=\sum_{c}\epsilon_{abc\,}\xi_{c}\,,}
\ea
\label{KillingA}
\ee
where $\mathbf{i}_{\xi_{a}}$ denotes the inner product. Since, in general, the exterior derivative and the Lie derivative commute, \textit{i.e.~}$[\rd,\cL_{\xi_{a}}]=0$, Eq.(\ref{Killing}) implies that there must be a one-form, $\lambda_{a}$, for each Killing vector, $\xi_{a}$, satisfying
\be
\cL_{\xi_{a}}B_{\scriptscriptstyle{(2)}}=-\rd{\lambda_{a}}\,.
\ee
Explicitly, we have
\be
\ba{lll}
\lambda_{1}=\dis{\frac{\cos\varphi}{\sin\vartheta}}\left[h\rd t+B(r)\rd r\right]\,,\quad&\quad~~\lambda_{2}=\dis{\frac{\sin\varphi}{\sin\vartheta}}\left[h\rd t+B(r)\rd r\right]\,,\quad&\quad~~\lambda_{3}=0\,.
\ea
\label{lambda}
\ee
It follows that the DFT-Killing equations hold:  both  the DFT-metric, $\cH_{AB}$, and  the DFT-dilaton are  annihilated by the generalized Lie derivative, \textit{c.f.~}\cite{Park:2015bza},
\be
\ba{ll}
\hcL{}_{V_{a}}\cH_{AB}=0\,,\quad~~&\quad~~\hcL_{V_{a}}\left(e^{-2d}\right)=0\,,
\ea
\ee
where the $\Off$ vectorial parameter is given by the $\so(3)$ angular momenta~(\ref{Killing}) and the  one-forms~(\ref{lambda}), 
\be
\ba{ll}
V^{A}_{a}=(\lambda_{a\mu},\xi_{a}^{\nu})\,,\qquad&\qquad
\big[V_{a}\,,\,V_{b}\big]_{\mathbf{C}}
=\sum_{c}\epsilon_{abc}V_{c}\,.
\ea
\ee
~\\

\noindent The nontrivial Christoffel symbols for the metric ansatz~(\ref{metric}) are exhaustively, 
\be
\ba{ll}
\Gamma^{t}_{tr}=\Gamma^{t}_{rt}
=\half{A^{\prime}}A^{-1}+{\phi^{\prime}}\,,\qquad&\qquad\quad
\Gamma^{r}_{tt}=\half{A^{\prime}}A+{\phi^{\prime}}A^{2}\,,\\
\Gamma^{r}_{rr}=-\half{A^{\prime}}A^{-1}+{\phi^{\prime}}\,,\qquad&\qquad\quad
\Gamma^{r}_{\vartheta\vartheta}=\half{A^{\prime}}A^{-1}C-\half{C^{\prime}}-C{\phi^{\prime}}\,,\\
\Gamma^{r}_{\varphi\varphi}
=\sin^{2}\vartheta\Gamma^{r}_{\vartheta\vartheta}\,,\qquad&\qquad\quad
\Gamma^{\vartheta}_{r\vartheta}=\Gamma^{\vartheta}_{\vartheta r}=-\half{A^{\prime}}A^{-1}
+\half{C^{\prime}}C^{-1}+{\phi^{\prime}}\,,\\
\Gamma^{\vartheta}_{\varphi\varphi}=-\sin\vartheta\cos\vartheta\,,\qquad&\qquad\quad
\Gamma^{\varphi}_{r\varphi}=\Gamma^{\varphi}_{\varphi r}=
-\half{A^{\prime}}A^{-1}+\half{C^{\prime}}C^{-1}+{\phi^{\prime}}\,,\\
\Gamma^{\varphi}_{\vartheta\varphi}=\Gamma^{\varphi}_{\varphi\vartheta}
=\cot\vartheta\,,\qquad&\qquad
\ea
\ee
where the prime denotes the radial derivative. \\
~\\

\noindent The Ricci curvature, $R_{\mu\nu}$, and  the second derivative, $\trd_{\mu}\partial_{\nu}\phi$, are automatically diagonal, such that the equation of motion of the string frame metric, \textit{i.e.}~(\ref{EOMg})   is  almost diagonal,
\be
\ba{rll}
R_{tt}+2\trd_{t}\partial_{t}\phi-\quarter H_{t\rho\sigma}H_{t}{}^{\rho\sigma}&=&\half {A^{\prime}}A\frac{\rd~}{\rd r}\ln({A^{\prime}}A^{-1}C)+{\phi^{\prime}}A^{2}\frac{\rd~}{\rd r}\ln({\phi^{\prime}}C)-\half h^{2}A^{2}C^{-2}e^{-4\phi}\,,\\
R_{rr}+2\trd_{r}\partial_{r}\phi-\quarter H_{r\rho\sigma}H_{r}{}^{\rho\sigma}&=&\half {A^{\prime}}A^{-1}\frac{\rd~}{\rd r}\ln({A^{\prime}}A^{-1}C)-\half{A^{\prime}}^{2}A^{-2}
-{C^{\prime\prime}}C^{-1}+\half{C^{\prime}}^{2}C^{-2}\\
{}&{}&-2{\phi^{\prime}}^{2}-{\phi^{\prime}}\frac{\rd~}{\rd r}\ln({\phi^{\prime}}C)-\half A^{2}B^{2}C^{-2}e^{-4\phi}\,,\\
R_{\vartheta\vartheta}+2\trd_{\vartheta}\partial_{\vartheta}\phi-\quarter H_{\vartheta\rho\sigma}H_{\vartheta}{}^{\rho\sigma}&=&
1+\half C({A^{\prime\prime}}A^{-1}-{C^{\prime\prime}}C^{-1})-\half{A^{\prime}}^{2}A^{-2}C
+\half{A^{\prime}}A^{-1}{C^{\prime}}\\
{}&{}&-{\phi^{\prime}}C\frac{\rd~}{\rd r}\ln({\phi^{\prime}}C)-\half(A^{2}B^{2}-h^{2})C^{-1}e^{-4\phi}\,,\\
R_{\varphi\varphi}+2\trd_{\varphi}\partial_{\varphi}\phi
-\quarter H_{\varphi\rho\sigma}H_{\varphi}{}^{\rho\sigma}&=&\sin^{2}\vartheta\left(R_{\vartheta\vartheta}+2\trd_{\vartheta}\partial_{\vartheta}\phi-\quarter H_{\vartheta\rho\sigma}H_{\vartheta}{}^{\rho\sigma}\right)\,.
\ea
\ee
The only  exception   is the following  off-diagonal component,
\be
R_{tr}+2\trd_{t}\partial_{r}\phi-\quarter H_{t\rho\sigma}H_{r}{}^{\rho\sigma}=-\quarter H_{t\rho\sigma}H_{r}{}^{\rho\sigma}=-\half\, hBe^{-4\phi}A^{2}C^{-2}\,.
\label{eorm}
\ee
This implies either $h=0\,$ or $\,B(r)=0$.  The remaining equations of motion~(\ref{EOMB2}), (\ref{EOMphi2}) become
\be
\rd\star\left(e^{-2\phi}H_{\scriptscriptstyle{(3)}}\right)=\rd \left(e^{-4\phi}A^{2}BC^{-1}\rd t+e^{-4\phi}hC^{-1}\rd r\right)=\frac{\rd~}{\rd r}\left(e^{-4\phi}A^{2}BC^{-1}\right)
\rd r\wedge\rd t=0\,,
\label{EOMB3}
\ee
and
\be
\Box\phi-2\partial_{\mu}\phi\partial^{\mu}\phi+
\textstyle{\frac{1}{12}}H_{\mu\nu\rho}H^{\mu\nu\rho}=
e^{-2\phi}{\phi^{\prime}}A\frac{\rd~}{\rd r}\ln({\phi^{\prime}}C)+\half e^{-6\phi}(A^{3}B^{2}C^{-2}-h^{2}AC^{-2})=0\,.
\ee
~\\

\noindent In this way,  all the equations of motion boil down to the following six relations:
\be
\ba{l}
e^{-4\phi}A^{2}BC^{-1}=q\,,\\
hq=0\,,\\
{C^{\prime\prime}}-2= q^{2} A^{-2}Ce^{4\phi}\,,\\
{A^{\prime}}A\,\frac{\rd~}{\rd r}\ln({A^{\prime}}A^{-1}C)=q^{2}e^{4\phi}\,,\\
{\phi^{\prime}}\frac{\rd~}{\rd r}\ln({\phi^{\prime}}C)=
\half h^{2}C^{-2}e^{-4\phi}-\half q^{2}A^{-2}e^{4\phi}\,,\\
4{\phi^{\prime}}^{2}+{A^{\prime}}^{2}A^{-2}
+{C^{\prime\prime}}C^{-1}-{C^{\prime}}^{2}C^{-2}+2C^{-1}
+h^{2}C^{-2}e^{-4\phi}=0\,,
\ea
\label{DEQ}
\ee
where $h,q$ are constants, associated with the electric and magnetic $H$-fluxes, see \eqref{Hflux}.  Apparently, either $h$ or $q$ must be trivial.  If ${h=0}$, ${q\neq0}$ and hence the $H$-flux were magnetic, it is easy to see that the above relations are inconsistent with  the asymptotically  flat smoothness  boundary conditions, (\ref{BC}) and (\ref{BC2}).  Henceforth we set  ${q\equiv0}$, ${B(r)\equiv0}$,  and focus on  electric $H$-flux solutions. The  differential equations above reduce to
\begin{eqnarray}
&&{C^{\prime\prime}}=2\,,\label{ddC}\\
&&\frac{\rd~}{\rd r}\left({A^{\prime}}A^{-1}C\right)=0\,,\label{AC}\\
&&{\phi^{\prime\prime}}+{\phi^{\prime}}{C^{\prime}}C^{-1}=\half h^{2}C^{-2}e^{-4\phi}\,,\label{ddotphi}\\
&&4{\phi^{\prime}}^{2}C^{2}+({B^{\prime}}A^{-1}C)^{2}
+4C-{C^{\prime}}^{2}
+h^{2}e^{-4\phi}=0\,.\label{dotphi2}
\end{eqnarray}
~\\

\noindent From (\ref{ddC}) and (\ref{AC}), imposing the boundary condition of (\ref{BC}), $A(r)$ and $C(r)$ are straightforwardly determined,
\be
\ba{ll}
A(r)=\left(\frac{r-c_{+}}{r-c_{-}}\right)^{\frac{a}{\,c_{+}-c_{-}}}\,,\quad&\quad
C(r)=(r-c_{+})(r-c_{-})\,,
\ea
\label{ACr}
\ee
where $a$ is the constant of the integral from (\ref{AC}), and $c_{+},c_{-}$ are two roots of the quadratic real polynomial,  $C(r)$.  Taking the radial derivative of (\ref{dotphi2}) gives nothing but  the second order differential equation~\eqref{ddotphi}, Thus, we only need to solve (\ref{dotphi2})  which becomes with the substitution of (\ref{ACr}),
\be
4\left[(r-c_{+})(r-c_{-}){\phi^{\prime}}\right]^{2}+a^{2}+h^{2}e^{-4\phi}=(c_{+}-c_{-})^{2}\,.
\label{dotphi22}
\ee
This result implies that --\,as the left hand side of the equality is positive\,-- the two roots, $c_{+}$, $c_{-}$, must be real, and further that eq.(\ref{dotphi22})  can be rewritten as   an integral relation,
\be
\dis{
\pm\int\frac{2\rd\phi}{\sqrt{(c_{+}-c_{-})^{2}-a^{2}-h^{2}e^{-4\phi}}}=\int\frac{\rd r}{(r-c_{+})(r-c_{-})}\,.}
\label{phirINT}
\ee
The left hand side integral gives
\be
\dis{
\pm\int\frac{2\rd\phi}{\sqrt{(c_{+}-c_{-})^{2}-a^{2}-h^{2}e^{-4\phi}}}= b^{-1}\ln\left(e^{2\phi}+\sqrt{e^{4\phi}-h^{2}b^{-2}}\right)\,+\,\mbox{constant}\,,
}
\label{phiintegral}
\ee
where we have absorbed the  sign factor, $\pm$, into   the newly introduced integration constant, $b$, satisfying
\be
a^{2}+b^{2}=(c_{+}-c_{-})^{2}\,.
\label{abc}
\ee
From   \eqref{dotphi22}, we note that $(c_{+}-c_{-})^{2}-a^{2}$ is positive and hence $b$ is a real number too. The right hand side of (\ref{phirINT}) gives
\be\dis{
\int\frac{\rd r}{(r-c_{+})(r-c_{-})}=\frac{1}{c_{+}-c_{-}}\ln\left(\frac{r-c_{+}}{r-c_{-}}\right)\,+\,\mbox{constant\,}^{\prime}\,.}
\label{rintegral}
\ee
~\\

\noindent Combining (\ref{phiintegral}) with (\ref{rintegral}) and   fixing the  integration constant from the boundary condition~(\ref{BC}), we obtain
\be
\textstyle{
e^{2\phi}=\half\left(1+\sqrt{1-h^{2}b^{-2}}\right)
\left(\frac{r-c_{+}}{r-c_{-}}\right)^{\frac{b}{\sqrt{a^{2}+b^{2}}}}+\half\left(1-\sqrt{1-h^{2}b^{-2}}\right)
\left(\frac{r-c_{+}}{r-c_{-}}\right)^{\frac{-b}{\sqrt{a^{2}+b^{2}}}}\,.
}
\label{phiSOL}
\ee
Thus, with four real  constants, $a,b,c,h$, and 
\be
\ba{lll}
c_{+}=c+\half\sqrt{a^{2}+b^{2}}\,,\quad&\quad
c_{-}=c-\half\sqrt{a^{2}+b^{2}}\,,\quad&\quad \gamma_{\pm}=\half\left(1\pm\sqrt{1-{h^{2}/b^{2}}}\right)\,,
\ea
\label{cpm}
\ee
the string frame metric takes the form,
\be{
\ba{ll}
\rd s^{2}\,=&-\left[\gamma_{+}
\left(\frac{r-c_{+}}{r-c_{-}}\right)^{\frac{a+b}{\sqrt{a^{2}+b^{2}}}}+\gamma_{-}\left(\frac{r-c_{+}}{r-c_{-}}\right)^{\frac{a-b}{\sqrt{a^{2}+b^{2}}}}\right]\rd t^{2}\\
{}&+
\left[\gamma_{+}\left(\frac{r-c_{+}}{r-c_{-}}\right)^{\frac{-a+b}{\sqrt{a^{2}+b^{2}}}}+\gamma_{-}\left(\frac{r-c_{+}}{r-c_{-}}\right)^{\frac{-a-b}{\sqrt{a^{2}+b^{2}}}}\right]\Big[\rd r^{2}+(r-c_{+})(r-c_{-})\rd\Omega^{2}\Big]\,,
\ea}
\label{metricSOL}
\ee
and the $B$-field  as  well as the $H$-flux are
\be
\ba{ll}
B_{\scriptscriptstyle{(2)}}=h\cos\vartheta\,\rd t\wedge\rd\varphi\,,\quad~&~\quad
H_{\scriptscriptstyle{(3)}}=h\sin\vartheta\,\rd t\wedge\rd\vartheta\wedge\rd\varphi\,.
\ea
\label{BSOL}
\ee
For the metric to be real  valued,  we must require  
\be
b^{2}\geq h^{2}\,.
\label{brange}
\ee
Namely, the presence of the electric $H$-flux (${h\neq 0}$)  induces the nontrivial string dilaton,  (${b\neq 0}$), see (\ref{phiSOL}).  The results of (\ref{phiSOL}), (\ref{metricSOL}) and (\ref{BSOL}) provide the most general form of  the  static, asymptotically flat and spherically symmetric vacuum  solutions to  $D=4$ Double Field Theory.\\

\noindent Up to the radial diffeomorphisms, we may set the free parameter, $c$, arbitrarily.  It is worth while then to rewrite the general solution in slightly different styles.

\begin{itemize}
\item  Firstly,   shifting the radial coordinate by constant, or putting $c\equiv\half\sqrt{a^{2}+b^{2}}$ for \eqref{cpm}, we may rewrite  the solution as (\ref{SOL2}):
\be
{\ba{cc}
\!\!\!\!\!\!\!\!\!\!\!e^{2\phi}=\gamma_{+}
\left(1-\frac{\sqrt{a^{2}+b^{2}}}{r}\right)^{\frac{b}{\sqrt{a^{2}+b^{2}}}}
+\gamma_{-}
\left(1-\frac{\sqrt{a^{2}+b^{2}}}{r}\right)^{\frac{-b}{\sqrt{a^{2}+b^{2}}}}\,,\quad&\quad
B_{\scriptscriptstyle{(2)}}=h\cos\vartheta\,\rd t\wedge\rd\varphi\,,\\
\multicolumn{2}{c}{\!\!\!\!\!\!\!\!\!\!\!
\rd s^{2}=e^{2\phi}\left[-\left(1-\frac{\sqrt{a^{2}+b^{2}}}{r}\right)^{\frac{a}{\sqrt{a^{2}+b^{2}}}}\rd t^{2}
+\left(1-\frac{\sqrt{a^{2}+b^{2}}}{r}\right)^{\frac{-a}{\sqrt{a^{2}+b^{2}}}}\left(\rd r^{2}+r\left(r-\sqrt{a^{2}+b^{2}}\right)
\rd\Omega^{2}\right)\right]\,,}
\ea}
\label{SOL20}
\ee
where the radial origin, $r=0$, corresponds to the coordinate singularity. \\
~\\

\item  Alternatively, if we choose $c\equiv\half\left(\frac{a-b}{a+b}\right)\sqrt{a^{2}+b^{2}}$ for \eqref{cpm}, 
the solution can be rewritten as \eqref{THESOL0} which we recall here:
\be
{\ba{cc}
e^{2\phi}=\gamma_{+}\left(\frac{r-\alpha}{r+\beta}\right)^{\frac{b}{\sqrt{a^{2}+b^{2}}}}+\gamma_{-}\left(\frac{r-\alpha}{r+\beta}\right)^{\frac{-b}{\sqrt{a^{2}+b^{2}}}}\,,\qquad&\qquad
B_{\scriptscriptstyle{(2)}}=h\cos\vartheta\,\rd t\wedge\rd\varphi\,,\\
\multicolumn{2}{c}{
\rd s^{2}=e^{2\phi}\left[-\left(\frac{r-\alpha}{r+\beta}\right)^{\frac{a}{\sqrt{a^{2}+b^{2}}}}\rd t^{2}
+\left(\frac{r-\alpha}{r+\beta}\right)^{\frac{-a}{\sqrt{a^{2}+b^{2}}}}\left(\rd r^{2}+(r-\alpha)(r+\beta)
\rd\Omega^{2}\right)\right]\,,}
\ea}
\label{SOL3}
\ee
where  
\be\textstyle{
\ba{lll}
\alpha=\frac{a}{a+b}\sqrt{a^{2}+b^{2}}\,,\quad&\quad
\beta=\frac{b}{a+b}\sqrt{a^{2}+b^{2}}\,,\quad&\quad
\gamma_{\pm}=\half\left(1\pm\sqrt{1-{h^{2}/b^{2}}}\right)\,.
\ea}
\ee
After all,  there are  three real parameters left, $a,b,h$, which satisfy $\,b^{2}\geq h^{2}$ and  have the dimension of length.\\

\end{itemize}
~\\

\noindent One side remark is that, from (\ref{ddotphi}), we may get the dual `axion', $\Phi$,  which has been a dark-matter candidate,  \textit{c.f.~}\cite{Kim:1979if,Shifman:1979if,Dine:1981rt,Zhitnitsky:1980tq,Sin:1992bg,Lee:1995af},
\[
\ba{ll}
\star\left( e^{-2\phi}H_{\scriptscriptstyle{(3)}}\right)
=he^{-4\phi}C^{-1}\rd r=2h^{-1}\rd ({\phi^{\prime}}C)=\rd\Phi\,,
\quad&\quad\Phi=2h^{-1}{\phi^{\prime}}C\,.
\ea
\]
We refer  readers to \cite{Duff:1994an} for the discussion on the related  S-duality.\\

\section{Derivation of the orbital velocity, Eq.(\ref{RvarphiV}) \label{APPGEO}}

For a planar orbital motion where the $\vartheta$ angle is fixed at $\vartheta\equiv\frac{\pi}{2}$, the geodesic equation  for  the other angle, $\varphi$,  reads
\be
{\frac{\rd^{2}\varphi}{\rd\tau^{2}}+2\Gamma^{\varphi}_{\varphi r}\frac{\rd\varphi}{\rd\tau}\frac{\rd r}{\rd\tau}=
\frac{\rd^{2}\varphi}{\rd\tau^{2}}+\frac{\rd\varphi}{\rd\tau}
\frac{\rd~}{\rd\tau}\ln\left(e^{2\phi}A^{-1}C\right)=0\,.}
\ee
This gives the conservation of the angular momentum,
\be
\ba{ll}
L_{\varphi}:= e^{2\phi}A^{-1}C\,\frac{\rd\varphi}{\rd\tau}\,,\qquad&\qquad
\frac{\rd L_{\varphi}}{\rd\tau}=0\,.
\ea
\ee
Further, if the orbital motion is circular, we have $\frac{\rd r}{\rd\tau}\equiv0$, and  the geodesic formula for  the radial coordinate,
\be
\small{
\frac{\rd^{2}r}{\rd\tau^{2}}+\Gamma^{r}_{tt}\left(\frac{\rd t}{\rd\tau}\right)^{2}+\Gamma^{r}_{\varphi\varphi}
\left(\frac{\rd\varphi}{\rd\tau}\right)^{2}=0\,,
}
\ee
determines the angular velocity,
\be
\small{\left(\frac{\rd\varphi}{\rd t}\right)^{2}=-\left(\frac{\rd g_{tt}}{\rd r}\right)\left(\frac{\rd g_{\varphi\varphi}}{\rd r}\right)^{-1}=\frac{\frac{\rd~}{\rd r}(Ae^{2\phi})}{\,\frac{\rd~}{\rd r}(CA^{-1}e^{2\phi})\,}\,.}
\label{varthetat}
\ee
Especially for a massless particle or   photon to be captured in a  circular orbit, and thus to form  a `photon sphere',  we further require  the `null' condition: directly from the metric ansatz~(\ref{metric}),
\be
\small{\left(\frac{\rd\varphi}{\rd t}\right)^{2}=A^{2}C^{-1}=\,-\,\frac{\,g_{tt}\,}{g_{\vartheta\vartheta}}\,.}
\ee
This relation must match with (\ref{varthetat}). Straightforward computation shows that,  the photon sphere is located at a radius, $r=\rphoton$, extremizing the angular velocity,
\be
\small{\left.
\frac{\rd~}{\rd r}\left(A^{2}C^{-1}\right)\right|_{r=\rphoton}=0\,.}
\label{photonsphere}
\ee
As it should be, the dilaton factor has disappeared above, thanks to the null property of the photon.\\
~\\

\noindent With the the `proper' radius,
\be
R:=\sqrt{g_{\vartheta\vartheta}(r)}=\sqrt{C(r)/A(r)}\,e^{\phi(r)}\,,
\ee
the angular part of the metric is properly normalized,
\be\textstyle{
\rd s^{2} =g_{tt}\rd t^{2}+g_{\scriptscriptstyle{RR}}\rd R^{2}+R^{2}\rd\Omega^{2}=-e^{2\phi}A\,\rd t^{2} +e^{2\phi}A^{-1}
\left(\frac{\rd R}{\rd r}\right)^{-2}\rd R^{2}+R^{2}\rd\Omega^{2}\,,}
\label{metric2}
\ee
and  in particular,   the circumference of a circle is $2\pi R$.\\
~\\

\noindent Explicitly for the  most general  spherically symmetric solution~(\ref{SOL3}),  
 the radius of the photon sphere is 
\be
\textstyle{
\rphoton=a+\half\left(\frac{a-b}{a+b}\right)\sqrt{a^{2}+b^{2}}\,\,;}
\ee
the angular velocity is 
\be
\left(\frac{\rd\varphi}{\rd t}\right)^{2}=\textstyle{{\frac{1}{
\left(r-\alpha\right)
\left(r+\beta\right)}}
\left(\frac{r-\alpha}{r+\beta}\right)^{\frac{2a}{\sqrt{a^{2}+b^{2}}}}
\left[\frac{\gamma_{+}(a+b)\left(\frac{r-\alpha}{r+\beta}\right)^{\frac{2b}{\sqrt{a^{2}+b^{2}}}}~+\,\gamma_{-}(a-b)
}{\gamma_{+}
\left(2r-\alpha+\beta-a+b\right)\left(\frac{r-\alpha}{r+\beta}\right)^{\frac{2b}{\sqrt{a^{2}+b^{2}}}}~
+\,\gamma_{-}\left(2r-\alpha+\beta-a-b\right)}
\right]\,;}
\label{varthetat2}
\ee
the proper radius is
\be
\textstyle{
R=\left[(r-\alpha)(r+\beta)\left(\gamma_{+}\left(\frac{r-\alpha}{r+\beta}\right)^{\frac{-a+b}{\sqrt{a^{2}+b^{2}}}}+
\gamma_{-}\left(\frac{r-\alpha}{r+\beta}\right)^{\frac{-a-b}{\sqrt{a^{2}+b^{2}}}}\right)\right]^{\frac{1}{2}}\,;}
\label{properR}
\ee
and  the orbital velocity  is
\be
\textstyle{
\!\!\!\Vo=\left|R\frac{\rd\varphi}{\rd t}\right|=\left[-\half R\frac{\,\rd g_{tt}}{\rd R}\,\right]^{\frac{1}{2}}=
\left[
\frac{\left(\gamma_{+}\left(\frac{r-\alpha}{r+\beta}\right)^{\frac{a+b}{\sqrt{a^{2}+b^{2}}}}+
\gamma_{-}\left(\frac{r-\alpha}{r+\beta}\right)^{\frac{a-b}{\sqrt{a^{2}+b^{2}}}}\right)\left(\gamma_{+}(a+b)\left(\frac{r-\alpha}{r+\beta}\right)^{\frac{2b}{\sqrt{a^{2}+b^{2}}}}+\gamma_{-}(a-b)\right)
}{\gamma_{+}
\left(2r-\alpha+\beta-a+b\right)\left(\frac{r-\alpha}{r+\beta}\right)^{\frac{2b}{\sqrt{a^{2}+b^{2}}}}
+\gamma_{-}\left(2r-\alpha+\beta-a-b\right)}
\right]^{\frac{1}{2}}
\,.}
\label{Vor}
\ee

\newpage
\section{Derivation of the global Noether charge, Eq.(\ref{observables}) \label{APPCHARGE}}

In \cite{Park:2015bza},  the general formula of the conserved global charge in DFT has been worked out, including  also  the   Yang-Mills sector~\cite{Jeon:2011kp}. For pure DFT,   the conserved global charge  reads 
\be
Q[X]=\oint_{\partial\cM}
\rmd^{{\scriptscriptstyle{2}}}x_{AB}~e^{-2d}\left(K^{AB}+2X^{[A}B^{B]}\right)\,,
\label{defQ}
\ee
where $\partial\cM$ denotes the  spatial infinity; $K^{AB}$ is a skew-symmetric Noether potential for DFT, 
\be
K^{AB}=4(\brP\na)^{[A}(PX)^{B]}-4(P\na)^{[A}(\brP X)^{B]}\,;
\label{Kpotential}
\ee
and  the second  term, $2X^{[A}B^{B]}$,  corresponds to  the  DFT extension of the counter two-form \textit{a la~}Wald~\cite{Wald:1993nt,Iyer:1994ys,Iyer:1995kg}. Explicitly,  $B^{A}$   is given  by
\be
B^{A}=2(P^{AC}P^{BD}-\brP^{AC}\brP^{BD})
\Gamma_{BCD}  =4(P-\brP)^{AB}\partial_{B}d -2\partial_{B}P^{AB}\,.
\label{defB}
\ee
The global charge~(\ref{defQ}) is conserved if $X^{A}$  meets
\be
\partial_{A}\partial_{[B}X_{C]}=0\,.
\ee 
For example,   a constant vector, corresponding to a rigid translational symmetry, satisfies this condition. \\

\noindent In terms of the conventional   field  variables, $\{g_{\mu\nu}, B_{\mu\nu}, \phi\}$, of the closed string massless sector,   the conserved global charge above reduces to
\be
 Q[X]=\int_{\partial\cM}\rmd^{{\scriptscriptstyle{D{-2}}}}x_{\mu\nu} \sqrt{-g}\,e^{-2\phi}\,
 \left(K^{\mu\nu}[X]+2X^{[\mu}B^{\nu]}\right)\,,
\ee
where now, with $X^{A}=(\zeta_{\mu}+B_{\mu\rho}\xi^{\rho}\,,\,\xi^{\nu})$,
\be
\ba{ll}
K^{\mu\nu}[X]= 2\, \xi^{[\mu;\nu]} - H^{\mu\nu\rho}\, \zeta_\rho \,,\quad&\quad
 B^{\mu}
  =2g^{\mu\nu}\left(2\partial_{\nu} \phi -\partial_{\nu} \ln\sqrt{-g}\,\right) - \partial_{\nu} g^{\mu\nu} \,.
 \ea
 \ee
~\\
\noindent Specifically for the  general  solution~(\ref{SOL3}), it is straightforward to compute  the conserved global charge for   the time translational symmetry:
\be
{
\cQ[\partial_{t}]=\frac{1}{4}\left[a+\left(\frac{a-b}{a+b}\right)\sqrt{a^{2}+b^{2}}\,\right]\,.}
\label{Q}
\ee
As known for Jordan (\textit{i.e.~}string) frame, \textit{e.g.~}\cite{Faraoni:2006fx}, this time translational global charge is not necessarily positive definite.   We speculate that only  $M_{\scriptscriptstyle{\infty}}$ in (\ref{observables}) ought to be non-negative.

\newpage



\begin{thebibliography}{99}


\bibitem{Rubin:1980zd}
  V.~C.~Rubin, N.~Thonnard and W.~K.~Ford, Jr.,
  ``Rotational properties of 21 SC galaxies with a large range of luminosities and radii, from NGC 4605 /R = 4kpc/ to UGC 2885 /R = 122 kpc/,''
  Astrophys.\ J.\  {\bf 238} (1980) 471.




\bibitem{Milgrom:1983ca}
  M.~Milgrom,
  ``A Modification of the Newtonian dynamics as a possible alternative to the hidden mass hypothesis,''
  Astrophys.\ J.\  {\bf 270} (1983) 365.




\bibitem{Berezhiani:2015bqa}
  L.~Berezhiani and J.~Khoury,
  ``Theory of dark matter superfluidity,''
  Phys.\ Rev.\ D {\bf 92} (2015) 103510.


\bibitem{Persic:1995ru}
  M.~Persic, P.~Salucci and F.~Stel,
  ``The Universal rotation curve of spiral galaxies: 1. The Dark matter connection,''
  Mon.\ Not.\ Roy.\ Astron.\ Soc.\  {\bf 281} (1996) 27
  [astro-ph/9506004].
  
 
  
  
  
\bibitem{Sofue:2000jx}
  Y.~Sofue and V.~Rubin,
  ``Rotation curves of spiral galaxies,''
  Ann.\ Rev.\ Astron.\ Astrophys.\  {\bf 39} (2001) 137
  [astro-ph/0010594].
  


\bibitem{Siegel:1993xq}
  W.~Siegel,
  ``Two vierbein formalism for string inspired axionic gravity,''
  Phys.\ Rev.\ D {\bf 47} (1993) 5453. 



\bibitem{Siegel:1993th}
  W.~Siegel,
  ``Superspace duality in low-energy superstrings,''
  Phys.\ Rev.\ D {\bf 48} (1993) 2826.



\bibitem{Hull:2009mi}
  C.~Hull and B.~Zwiebach,
  ``Double Field Theory,''
  JHEP {\bf 0909} (2009) 099.



\bibitem{Hull:2009zb}
  C.~Hull and B.~Zwiebach,
  ``The Gauge algebra of double field theory and Courant brackets,''
  JHEP {\bf 0909} (2009) 090.



\bibitem{Hohm:2010jy}
  O.~Hohm, C.~Hull and B.~Zwiebach,
  ``Background independent action for double field theory,''
  JHEP {\bf 1007} (2010) 016.



\bibitem{Hohm:2010pp}
  O.~Hohm, C.~Hull and B.~Zwiebach,
  ``Generalized metric formulation of double field theory,''
  JHEP {\bf 1008} (2010) 008.



\bibitem{Jeon:2010rw}
  I.~Jeon, K.~Lee and J.~H.~Park,
  ``Differential geometry with a projection: Application to double field theory,''
  JHEP {\bf 1104} (2011) 014.



\bibitem{Hohm:2010xe}
  O.~Hohm and S.~K.~Kwak,
  ``Frame-like Geometry of Double Field Theory,''
  J.\ Phys.\ A {\bf 44} (2011) 085404.



\bibitem{Jeon:2011cn}
  I.~Jeon, K.~Lee and J.~H.~Park,
  ``Stringy differential geometry, beyond Riemann,''
  Phys.\ Rev.\ D {\bf 84} (2011) 044022.



\bibitem{Jeon:2011vx}
  I.~Jeon, K.~Lee and J.~H.~Park,
  ``Incorporation of fermions into double field theory,''
  JHEP {\bf 1111} (2011) 025.



\bibitem{Hohm:2011si}
  O.~Hohm and B.~Zwiebach,
  ``On the Riemann Tensor in Double Field Theory,''
  JHEP {\bf 1205} (2012) 126.


\bibitem{Jeon:2012kd}
  I.~Jeon, K.~Lee and J.~H.~Park,
  ``Ramond-Ramond Cohomology and O(D,D) T-duality,''
  JHEP {\bf 1209} (2012) 079.


\bibitem{Berman:2013uda}
  D.~S.~Berman, C.~D.~A.~Blair, E.~Malek and M.~J.~Perry,
  ``The $O_{D,D}$ geometry of string theory,''
  Int.\ J.\ Mod.\ Phys.\ A {\bf 29} (2014) 1450080.

\bibitem{Cederwall:2014kxa}
  M.~Cederwall,
  ``The geometry behind double geometry,''
  JHEP {\bf 1409} (2014) 070.


\bibitem{Hitchin:2004ut}
  N.~Hitchin,
  ``Generalized Calabi-Yau manifolds,''
  Quart.\ J.\ Math.\  {\bf 54} (2003) 281.





\bibitem{Coimbra:2011nw}
  A.~Coimbra, C.~Strickland-Constable and D.~Waldram,
  ``Supergravity as Generalised Geometry I: Type II Theories,''
  JHEP {\bf 1111} (2011) 091.




  
\bibitem{Aldazabal:2013sca}
  G.~Aldazabal, D.~Marques and C.~Nunez,
  ``Double Field Theory: A Pedagogical Review,''
  Class.\ Quant.\ Grav.\  {\bf 30} (2013) 163001
  [arXiv:1305.1907 [hep-th]].



\bibitem{Berman:2013eva}
  D.~S.~Berman and D.~Thompson,
  ``Duality Symmetric String and M-Theory,''
  Phys.\,Rept.\,{\bf 566}\,(2014)\,1.



\bibitem{Hohm:2013bwa}
  O.~Hohm, D.~L{u}st and B.~Zwiebach,
  ``The Spacetime of Double Field Theory: Review, Remarks, and Outlook,''
  Fortsch.\ Phys.\  {\bf 61} (2013) 926
  [arXiv:1309.2977 [hep-th]].



  

\bibitem{Duff:1986ne}
  M.~J.~Duff,
  ``Hidden String Symmetries?,''
  Phys.\ Lett.\ B {\bf 173} (1986) 289.



\bibitem{Choi:2015bga}
  K.~S.~Choi and J.~H.~Park,
  ``Standard Model as a Double Field Theory,''
  Phys.\ Rev.\ Lett.\  {\bf 115} (2015) no.17,  171603.
  
  





\bibitem{Jeon:2011kp}
  I.~Jeon, K.~Lee and J.~H.~Park,
  ``Double field formulation of Yang-Mills theory,''
  Phys.\ Lett.\ B {\bf 701} (2011) 260
  [arXiv:1102.0419 [hep-th]].
  
  
  


\bibitem{Jeon:2011sq}
  I.~Jeon, K.~Lee and J.~H.~Park,
  ``Supersymmetric Double Field Theory: Stringy Reformulation of Supergravity,''
  Phys.\ Rev.\ D {\bf 85} (2012) 081501
   Erratum: [Phys.\ Rev.\ D {\bf 86} (2012) 089903(E)].



\bibitem{Jeon:2012hp}
  I.~Jeon, K.~Lee, J.~H.~Park and Y.~Suh,
  ``Stringy Unification of Type IIA and IIB Supergravities under N=2 D=10 Supersymmetric Double Field Theory,''
  Phys.\ Lett.\ B {\bf 723} (2013) 245. 



\bibitem{Cho:2015lha}
  W.~Cho, J.~J.~Fernández-Melgarejo, I.~Jeon and J.~H.~Park,
  ``Supersymmetric gauged double field theory: systematic derivation by virtue of twist,''
  JHEP {\bf 1508} (2015) 084. 



\bibitem{Hohm:2011zr}
  O.~Hohm, S.~K.~Kwak and B.~Zwiebach,
  ``Unification of Type II Strings and T-duality,''
  Phys.\ Rev.\ Lett.\  {\bf 107} (2011) 171603. 






\bibitem{Burgess:1994kq}
  C.~P.~Burgess, R.~C.~Myers and F.~Quevedo,
  ``On spherically symmetric string solutions in four-dimensions,''
  Nucl.\ Phys.\ B {\bf 442} (1995) 75
  doi:10.1016/S0550-3213(95)00090-9
  [hep-th/9410142].




\bibitem{Fisher:1948yn}
  I.~Z.~Fisher,
  ``Scalar mesostatic field with regard for gravitational effects,''
  Zh.\ Eksp.\ Teor.\ Fiz.\  {\bf 18} (1948) 636.



\bibitem{Janis:1968zz}
  A.~I.~Janis, E.~T.~Newman and J.~Winicour,
  ``Reality of the Schwarzschild Singularity,''
  Phys.\ Rev.\ Lett.\  {\bf 20} (1968) 878.






\bibitem{Virbhadra:2002ju}
  K.~S.~Virbhadra and G.~F.~R.~Ellis,
  ``Gravitational lensing by naked singularities,''
  Phys.\ Rev.\ D {\bf 65} (2002) 103004.





\bibitem{Faraoni:1998qx}
  V.~Faraoni, E.~Gunzig and P.~Nardone,
  ``Conformal transformations in classical gravitational theories and in cosmology,''
  Fund.\ Cosmic Phys.\  {\bf 20} (1999) 121.





\bibitem{Park:2013mpa}
  J.{~H.}~Park,
  ``Comments on double field theory and diffeomorphisms,''
  JHEP {\bf 1306} (2013) 098.



\bibitem{Lee:2013hma}
  K.~Lee and J.~H.~Park,
  ``Covariant action for a string in "doubled yet gauged" spacetime,''
  Nucl.\ Phys.\ B {\bf 880} (2014) 134. 




\bibitem{Bergshoeff:2011se}
  E.~A.~Bergshoeff, T.~Ortin and F.~Riccioni,
  ``Defect Branes,''
  Nucl.\ Phys.\ B {\bf 856} (2012) 210
  doi:10.1016/j.nuclphysb.2011.10.037
  [arXiv:1109.4484 [hep-th]].


  

\bibitem{Malek:2013sp}
  E.~Malek,
  ``Timelike U-dualities in Generalised Geometry,''
  JHEP {\bf 1311} (2013) 185.

















\bibitem{Ko:2015rha}
  S.~M.~Ko, C.~Melby-Thompson, R.~Meyer and J.~H.~Park, 
  ``Dynamics of Perturbations in Double Field Theory \& Non-Relativistic String Theory,''
  JHEP {\bf 1512} (2015) 144
  [arXiv:1508.01121 [hep-th]].
  
  
  
  
  



\bibitem{Park:2015bza}
  J.~H.~Park, S.~J.~Rey, W.~Rim and Y.~Sakatani,
  ``O(D, D) covariant Noether currents and global charges in double field theory,''
  JHEP {\bf 1511} (2015) 131.
  
  
  

\bibitem{Hernquist:1990be}
  L.~Hernquist,
  ``An Analytical Model for Spherical Galaxies and Bulges,''
  Astrophys.\ J.\  {\bf 356} (1990) 359.




\bibitem{Wald:1993nt}
  R.~M.~Wald,
 ``Black hole entropy is the Noether charge,'' 
  Phys.\ Rev.\ D {\bf 48} (1993) 3427
  [gr-qc/9307038].


\bibitem{Iyer:1994ys}
  V.~Iyer and R.~M.~Wald, 
``Some properties of Noether charge and a proposal for dynamical black hole entropy,'' 
  Phys.\ Rev.\ D {\bf 50} (1994) 846
  [gr-qc/9403028].


\bibitem{Iyer:1995kg}
  V.~Iyer and R.~M.~Wald, 
``A Comparison of Noether charge and Euclidean methods for computing the entropy of stationary black holes,'' 
  Phys.\ Rev.\ D {\bf 52} (1995) 4430
  [gr-qc/9503052].
  
  



\bibitem{Blair:2015eba}
  C.~D.~A.~Blair,
  ``Conserved Currents of Double Field Theory,''
  JHEP {\bf 1604} (2016) 180.
  
  
  
\bibitem{Kim:1979if}
  J.~E.~Kim,
  ``Weak Interaction Singlet and Strong CP Invariance,''
  Phys.\ Rev.\ Lett.\  {\bf 43} (1979) 103.
  
\bibitem{Shifman:1979if}
  M.~A.~Shifman, A.~I.~Vainshtein and V.~I.~Zakharov,
  ``Can Confinement Ensure Natural CP Invariance of Strong Interactions?,''
  Nucl.\ Phys.\ B {\bf 166} (1980) 493.
  
  
\bibitem{Dine:1981rt}
  M.~Dine, W.~Fischler and M.~Srednicki,
  ``A Simple Solution to the Strong CP Problem with a Harmless Axion,''
  Phys.\ Lett.\ B {\bf 104} (1981) 199.
  
\bibitem{Zhitnitsky:1980tq}
  A.~R.~Zhitnitsky,
  ``On Possible Suppression of the Axion Hadron Interactions. (In Russian),''
  Sov.\ J.\ Nucl.\ Phys.\  {\bf 31} (1980) 260
   [Yad.\ Fiz.\  {\bf 31} (1980) 497].
 
     



\bibitem{Sin:1992bg}
  S.~J.~Sin,
  ``Late time cosmological phase transition and galactic halo as Bose liquid,''
  Phys.\ Rev.\ D {\bf 50} (1994) 3650
  doi:10.1103/PhysRevD.50.3650
  [hep-ph/9205208].



\bibitem{Lee:1995af}
  J.~Lee and I.~Koh,
  ``Galactic halos as boson stars,''
  Phys.\ Rev.\ D {\bf 53} (1996) 2236
  doi:10.1103/PhysRevD.53.2236
  [hep-ph/9507385].


  
  
  
  
\bibitem{Clowe:2006eq}
  D.~Clowe, M.~Bradac, A.~H.~Gonzalez, M.~Markevitch, S.~W.~Randall, C.~Jones and D.~Zaritsky,
  ``A direct empirical proof of the existence of dark matter,''
  Astrophys.\ J.\  {\bf 648} (2006) L109
  doi:10.1086/508162
  [astro-ph/0608407].



\bibitem{Duff:1994an}
  M.~J.~Duff, R.~R.~Khuri and J.~X.~Lu,
  ``String solitons,''
  Phys.\ Rept.\  {\bf 259} (1995) 213
  [hep-th/9412184].
  

\bibitem{Faraoni:2006fx}
  V.~Faraoni and S.~Nadeau,
  ``The (pseudo)issue of the conformal frame revisited,''
  Phys.\ Rev.\ D {\bf 75} (2007) 023501
  doi:10.1103/PhysRevD.75.023501
  [gr-qc/0612075].







\end{thebibliography}

\end{document}